\documentclass{article}
\usepackage{graphicx} %
\usepackage{xcolor}
\usepackage{biblatex}
\usepackage{fullpage}
\usepackage{amsmath}
\usepackage{tabularray}
\usepackage{float}
\usepackage{sidecap}
\usepackage{authblk}

\newcommand{\drugs}{DRUGS}

\bibliography{references}

\title{Investigating the Behavior of Diffusion Models for Accelerating Electronic Structure Calculations}
\author[1]{Daniel Rothchild}
\author[2,3]{Andrew S. Rosen}
\author[4,3]{Eric Taw}
\author[5]{Connie Robinson}
\author[1]{Joseph E. Gonzalez}
\author[4,1]{Aditi S. Krishnapriyan}
\affil[1]{Department of Electrical Engineering and Computer Science, University of California, Berkeley}
\affil[2]{Department of Materials Science and Engineering, University of California, Berkeley}
\affil[3]{Materials Science Division, Lawrence Berkeley National Laboratory}
\affil[4]{Department of Chemical and Biomolecular Engineering, University of California, Berkeley}
\affil[5]{Department of Chemistry, University of California, Berkeley}
\date{November 2023}

\begin{document}

\maketitle

\begin{abstract}
    We present an investigation into diffusion models for molecular generation, with the aim of better understanding how their predictions compare to the results of physics-based calculations.
    The investigation into these models is driven by their potential to significantly accelerate electronic structure calculations using machine learning, without requiring expensive first-principles datasets for training interatomic potentials.
    We find that the inference process of a popular diffusion model for \emph{de novo} molecular generation is divided into an exploration phase, where the model chooses the atomic species, and a relaxation phase, where it adjusts the atomic coordinates to find a low-energy geometry.
    As training proceeds, we show that the model initially learns about the first-order structure of the potential energy surface, and then later learns about higher-order structure. 
    We also find that the relaxation phase of the diffusion model can be re-purposed to sample the Boltzmann distribution over conformations and to carry out structure relaxations.
    For structure relaxations, the model finds geometries with $\sim10\times$ lower energy than those produced by a classical force field for small organic molecules. Initializing a density functional theory (DFT) relaxation at the diffusion-produced structures yields a $>2\times$ speedup to the DFT relaxation when compared to initializing at structures relaxed with a classical force field.
\end{abstract}

\section{Introduction}
The potential energy surface (PES) is fundamental for understanding the behavior of atomic systems.
Finding stable geometries, estimating reaction rates, and predicting transition states all require an understanding of the shape of the PES.
Accurate first-principles methods for calculating points on the PES are computationally expensive or even entirely infeasible for larger systems, so a number of cheaper alternatives have been developed, such as classical force fields, semi-empirical tight-binding methods, and machine learning methods.

The vast majority of machine learning methods aimed at understanding the shape of the PES are learned interatomic potentials, and state-of-the-art interatomic potentials currently consist largely of neural-network interatomic potentials (NNIPs) \cite{deringer2019machine, friederich2021machine}.
NNIPs are trained through supervised learning, and the most common approach is to use a dataset of atomic geometries that are annotated with energies and forces. These energies and forces are typically derived from a more expensive physics-based method.
The result is a machine learning model that can predict, given a particular geometry of atoms, the energy of that atomic configuration and the forces experienced by each atom.

While machine-learned interatomic potentials are powerful tools for understanding the shape of the PES, they suffer from the major limitation that they require a supervised dataset of atomic geometries that have been annotated with energies and forces.
Relying on datasets based on first-principles physics-based calculations is challenging for several reasons:

\begin{itemize}
\item \textbf{Computational expense.} It is extremely computationally expensive to create this sort of dataset. For example, the OC20 dataset includes 1.3M relaxations, each of which was allowed up to 1728 core-hours of compute time \cite{chanussot2021open}.

\item \textbf{Lack of inclusion of all geometries.} The space of off-equilibrium geometries is combinatorially large, and there is no definitive way to choose which geometries to include in the dataset. If important types of geometries are excluded, models trained on the dataset may generalize poorly.

\item \textbf{Limited by the level of theory.} Because of how expensive it is to produce the dataset, we are limited in the level of theory that can be used for generating the energies and forces.
    For example, authors often generate geometries using tight-binding calculations, which are significantly less accurate than density functional theory (DFT) \cite{axelrod2022geom, tmqm}.
    Generating a labeled dataset of energies and forces based on experimental data is not feasible, so learned interatomic potentials will always be limited by the training set's level of theory.
\end{itemize}

To address these concerns, we investigate an alternative approach to understand the PES, which requires only ground-state (i.e., lowest energy) geometries as training data, and which does not involve explicitly learning an interatomic potential.
Recently, several authors have trained non-NNIP machine learning models to generate 3D atomic geometries, either from scratch or based on a molecular graph, by training a self-supervised model on a dataset of known 3D geometries \cite{jing2022torsional, hoogeboom2022equivariant, ganea2021geomol, reidenbach2023coarsenconf}.
In the self-supervised learning setting, models are trained on data without labels, which in this case means 3D atomic geometries that have not been annotated with energies or forces.
With no labels to learn from, models are instead trained by corrupting the training data with noise and then asking the model to predict the original, de-noised data.
The intended use of these models is to generate geometries (either from scratch or for conformer generation), rather than to reason more generally about the PES. However, they are undoubtedly learning about the PES, given that they are able to generate structures that lie near local energy minima.
To our knowledge, there has not been a comprehensive investigation into the degree to which these models acquire insights about the PES in the vicinity of the local minima, as opposed to solely learning point estimates of where the local minima lie. Existing evaluations only look at the generated geometries themselves to measure their quality.

The reason these models are worth exploring in more detail is that they provide a path forward to accelerate electronic structure calculations, while avoiding the challenges of training NNIPs described above.
Instead of training on a dataset of structures with their corresponding energies and forces, these models train in a self-supervised fashion on un-annotated ground-state geometries.
Training in this manner offers a number of advantages:

\begin{itemize}
    \item \textbf{Computational savings.} It is computationally cheaper to generate a dataset of only ground state configurations than to generate a training dataset for NNIPs, since we can initialize calculations at a higher level of theory with the best guess from a lower level of theory. In contrast, NNIP training data must include many off-equilibrium structures at a high level of theory so that they learn to generalize beyond the immediate neighborhood of the ground state. For example, in order to ensure the dataset is not ``biased toward structures with lower energies'', the Open Catalyst 2020 (OC20) dataset specifically includes DFT calculations on geometries ``with higher forces and greater configurational diversity'' than strictly necessary to identify the ground state. \cite{chanussot2021open}.
    \item \textbf{No need to select atomic geometries.} There is no need to choose which off-equilibrium geometries should be included, since a self-supervised dataset includes only ground state configurations. 
    \item \textbf{Can learn from experimental data.} Self-supervised methods offer a distinct advantage over NNIPs in that, in principle, they can be trained on experimental data. Training ML models on experimentally measured geometries is challenging, and we do not pursue experimental structures in this manuscript, but we believe that even the possibility of leveraging experimental data for training is a major advantage for self-supervised methods, when compared to NNIPs.
\end{itemize}

In this manuscript, we choose one of these models trained with self-supervised methods---an E(3) Equivariant Diffusion Model (EDM) \cite{hoogeboom2022equivariant}---and we probe the model's understanding of the PES 1) by inspecting more closely the model's predictions when generating structures as intended, 2) by examining the predictions when applied to downstream applications that it was not trained on, and 3) by comparing the predictions with the results of physics-based calculations (\S \ref{sec:investigation}).
Following this analysis, we investigate a practical approach to use EDM to accelerate structure relaxations of atomic systems (\S \ref{sec:accelerate}).

Our goal is not to train a better interatomic potential; prior work has already investigated denoising as a way to improve supervised learning, including for NNIPs \cite{zaidi2023pre, godwin2022simple, liu2022molecular}.
We are also not proposing a new way to train self-supervised models on chemical systems, rather opting for an off-the-shelf EDM model.
Instead, our objectives are: first, to understand, from a scientific standpoint, what these models are learning about the PES using only a denoising objective;
and second, to propose a practical way to use models trained in this manner to accelerate electronic structure calculations.

To summarize, we make the following contributions.
    We undertake a study of a pretrained EDM model, finding that its inference procedure can be roughly divided into an ``exploration'' regime and a ``relaxation'' regime (\S \ref{sec:edm_study}).
    For small organic molecules, we demonstrate that the relaxation regime of the EDM model finds geometries with significantly lower energies than those found using a classical force field (\S \ref{sec:relaxations}).
    When ``relaxing'' structures, EDM's predictions for how to de-noise the atomic positions preferentially follow the forces to the ground state early in training, and preferentially move straight towards the ground state later in training (\S \ref{sec:forces}).
    We use EDM to sample from a molecule's Boltzmann distribution over conformations, establishing a correspondence between diffusion steps and temperature (\S \ref{sec:boltzmann}).
    We re-purpose EDM to accelerate DFT structure relaxations, and we find that it can significantly speed up these calculations by proposing better initial geometries (\S \ref{sec:speedup}).
    We attain a small speedup to structure relaxations on a dataset of larger drug-like molecules; for these more complex PESs, EDM's predictions align better with the ground-truth forces than with the direct path to the ground state (\S \ref{sec:geom}).

\section{Related Work}
There is a long history of empirical force fields that sacrifice accuracy on molecular mechanics tasks to achieve computational efficiency.
Simple potentials like the Lennard-Jones \cite{jones1924determination}, Mie \cite{mie1903kinetischen}, and Kihara \cite{kihara1951second} potentials require only a few parameters to be tuned to match experiment or \emph{ab initio} simulations.
More complex potentials, such as
the Merck Molecular force field (MMFF) \cite{halgren1996merck},
the Universal Force Field (UFF) \cite{rappe1992uff},
Assisted Model Building with Energy Refinement (AMBER) \cite{weiner1984new, weiner1986all, cornell1995second},
Generalized AMBER force field (GAFF) \cite{wang2004development},
the ``Molecular Mechanics'' force fields (MM1 through MM4) \cite{allinger1976calculation, allinger1977conformational, allinger1989molecular, allinger1996improved},
the ``Chemistry at Harvard Macromolecular Mechanics'' force field (CHARMM) \cite{mackerell1998all},
and others model more complex interactions explicitly using many parameters, which must be fit to data.

More recently, neural network interatomic potentials (NNIPs) have taken this trend to the next level, introducing models that predict the energy of atomic configurations using millions of parameters, which must be trained using techniques from machine learning.
These NNIPs tend to be based on graph neural network architectures \cite{schutt2017schnet, unke2019physnet, gasteiger2020directional, gasteiger2021gemnet}, and many NNIPs are equivariant, meaning that rotating the input geometry leads to a deterministic and easy-to-calculate transformation of the output \cite{schutt2021equivariant, batzner2022nequip, qiao2022informing, brandstetter2022geometric, tholke2022equivariant, haghighatlari2022newtonnet,  musaelian2023learning}.
Energies predicted by an equivariant neural network are scalar quantities, so they are invariant to rotations of the input geometry. 
Predicted forces are vector quantities, subject to any rotation that is applied to the input geometry.
See \textcite{geiger2022e3nn} for a more in-depth primer on equivariance in neural networks.

All of these approaches---from the two-parameter Lennard-Jones potential to the largest GemNet \cite{gasteiger2021gemnet} model with millions of parameters---follow the traditional paradigm of predicting energies and forces.
These predicted energies and forces are then used to carry out structure relaxations, molecular dynamics simulations, Markov chain Monte Carlo simulations, etc.
In contrast, in this work we propose executing these tasks by training a machine learning model using a denoising objective on ground-state geometries, with no energy or force data in the training set.

Prior work has used denoising objectives in this domain for other purposes.
\textcite{hoogeboom2022equivariant}, whose model we use in this work, trains an equivariant denoising diffusion model to generate molecules from scratch.
\textcite{godwin2022simple} use a denoising objective as a regularization term for property prediction and one-shot structure relaxations.
\textcite{zaidi2023pre} is most similar to this work, as they use a similar denoising objective to pretrain a graph neural network for molecular property prediction.
They do not investigate the pretrained models themselves, focusing instead on using evaluating them only as starting points for fine-tuning a property prediction model. 
The success of their method motivates further study into what is learned via the denoising pretraining step.

\section{Background}
\label{sec:background}
Denoising diffusion models were recently popularized on the task of natural image generation \cite{ho2020denoising}.
The following offers a very high level and intuitive explanation of these models that leaves out important details. 
For a complete treatment, please refer to \cite{ho2020denoising} and \cite{hoogeboom2022equivariant}.

To train a diffusion model to generate images, the model is repeatedly given images from the training set that have been corrupted with Gaussian noise, and it is tasked with predicting what noise was added.
During training, the inputs to the model are corrupted with different amounts of noise. 
Sometimes, it faces images with only a small amount of noise, while at other times, the noise is so pronounced that the original image is nearly unrecognizable. 
Each time, the model is tasked with predicting what noise was added.
The amount of noise added is controlled by the ``diffusion step'', $n$, which ranges from $n=1$ (low noise) to $n=N$ (high noise), and the model takes $n$ as an input, in addition to the corrupted image.\footnote{The diffusion step is usually referred to as $t\in[1,T]$, but we reserve $T$ for temperature and use $n\in[1,N]$ for the diffusion step instead.}
During training, every iteration samples a random diffusion step uniformly from $n=1$ to $N$.
Consequently, the model learns to make any image less noisy---whether it is already very noisy or only a bit noisy.
To generate a new image during inference, the model is first given completely random Gaussian noise as the image, and $n=N$.
The noise predicted by the model is subtracted from the image (after appropriate scaling), and the result is fed back into the model with $n=N-1$.
This process continues until reaching $n=1$, at which point the original ``image'' has been completely denoised to obtain a purely generated image.

Recently, \textcite{hoogeboom2022equivariant} extended this technique to generating small organic molecules instead of natural images.
Intuitively, their model works the same way as is described above.
However, instead of generating images, the model generates molecules: the 3D coordinates of each atom, the chemical species of each atom, and the formal charge on each atom.
The 3D coordinates are represented simply as the usual scalar-valued $x$, $y$, and $z$ coordinates, and the atomic charges are also represented as scalars.
To represent the chemical species, the authors assign each atom in the training set a vector, where all values in the vector are zero except the entry corresponding to the true atom type, which is set to one (\emph{i.e.}, a one-hot vector).
At train time, the model receives these three quantities, each corrupted with Gaussian noise.
At test time, the model iteratively denoises what was originally an entirely random input until a plausible molecule emerges.
In order to respect translational symmetry, EDM translates all structures so that the center of mass is at the origin.
In order to respect rotational symmetry, EDM uses an equivariant neural network called EGNN to predict what noise was added \cite{satorras2021n}.

\section{Investigation Into Equivariant Diffusion Models}
\label{sec:investigation}
\subsection{Physical Intuition Behind EDM}
\label{sec:edm_study}
As a first step, and to motivate the subsequent discussion in this manuscript, we investigate in more detail how a trained EDM model generates 3D geometries.
We look at how structures evolve over the course of the diffusion process, with the aim of gaining a physical understanding of the model's behavior that will inform what other chemical tasks we might be able to carry out besides the intended \emph{de-novo} molecular generation.

\paragraph{Methods.}
As discussed in \S \ref{sec:background}, to carry out inference with an EDM model, first, we sample standard normal distributions for the atomic coordinates, for the values of the one-hot atomic species vector, and for the atomic charges. 
To ensure translation invariance, we subtract out the center of mass from the 3D coordinates, as described above.
Then, for each of $N$ steps, from $n=N$ down to $n=1$, we use an equivariant neural network to predict how the atomic coordinates, atomic species distribution, and charges should change, and we adjust the sampled molecule accordingly.

We train the EDM model on Quantum Machines 9 (QM9), a dataset of small molecules with up to nine heavy atoms among carbon, nitrogen, oxygen, and fluorine \cite{ruddigkeit2012enumeration, ramakrishnan2014quantum}, using the same hyperparameters used by \textcite{hoogeboom2022equivariant}.
We train the model for 6200 epochs and choose the model with the best validation loss, which occurs at epoch 5150.
In comparison, \textcite{hoogeboom2022equivariant} train the same model for 1100 epochs.
We find that the validation loss improves consistently until epoch $\sim$3000, after which it mostly levels off.

To match QM9, we carry out all DFT calculations at the B3LYP/6-31G(2df,p) level of theory \cite{becke1992density, stephens1994ab, ditchfield1971self, hehre1972self,krishnan1980self}.
We use Psi4 \cite{smith2020psi4} version 1.8, and relaxations are carried out with the Atomic Simulation Environment (ASE) \cite{ase-paper} version 3.22.1 using the BFGS algorithm for geometry optimizations with \texttt{fmax} of 0.03 eV/\AA.

\begin{figure}
    \centering
    \includegraphics[width=0.49\textwidth]{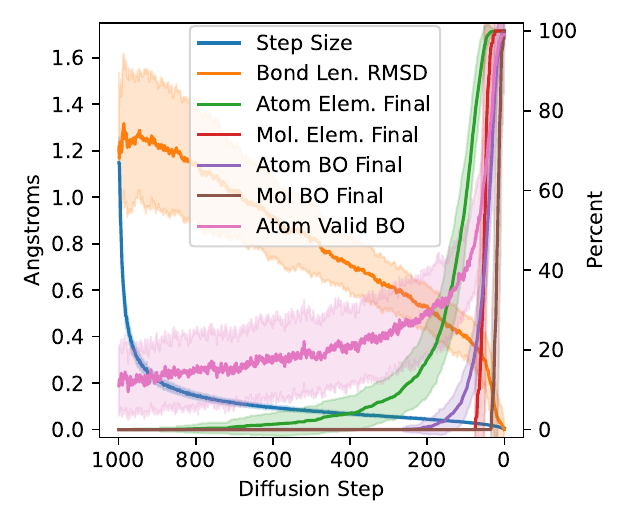}
    \includegraphics[width=0.49\textwidth]{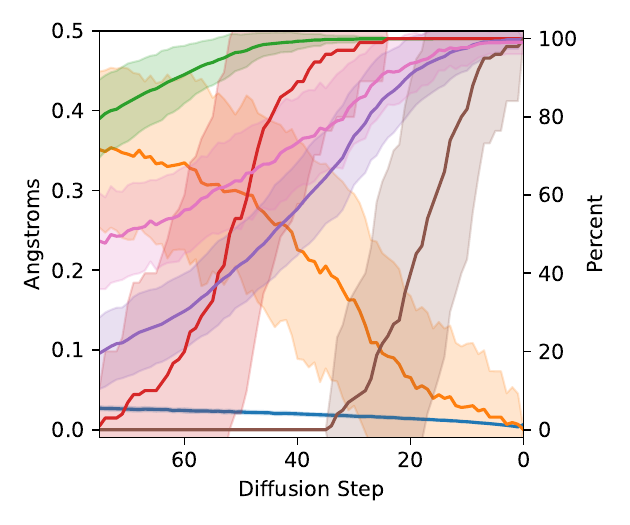}
    \caption{
    \textbf{Examining EDM Inference.}
    Left: $\ell_2$ norm of the step size taken at each step in diffusion (``Step Size'', \AA);
    root-mean-squared deviation between interatomic distances of the current structure compared to the final structure, considering only distances between atoms bonded in the final structure (``Bond Len. RMSD'', \AA);
    fraction of atoms whose chemical species is finalized (``Atom Elem. Final'');
    fraction of molecules where every atom's chemical species is finalized (``Mol. Elem. Final'');
    fraction of atoms that have the same number of bonds as they do in the final structure (``Atom BO Final'');
    fraction of molecules where all bond orders have been finalized (``Mol. BO Final'');
    fraction of atoms that have a valid number of bonds (e.g. 4 for carbon, ``Atom Valid BO'').
    Solid lines are the average of these quantities across 100 generated molecules, with shaded regions representing one standard deviation above/below.
    Right: Zoom of figure on the left.
    }
    \label{fig:diffusion_progression}
\end{figure}

\paragraph{Results.} At the beginning of the inference process, the atomic species and 3D positions are completely scrambled, but by the end, the model has decided on which atomic species to use and where to place them.
This progression is shown in Figure \ref{fig:diffusion_progression}, which plots as a function of diffusion steps the fraction of atomic species that have been finalized, the fraction of atoms that have the correct valence, and how close the interatomic distances are to their final values.

We first consider the plot of what fraction of molecules have all atomic species finalized (``Mol. Elem. Final'').
Here, it is clear that the diffusion process can be divided into two regimes: an ``exploration'' regime, from diffusion step 1000 to $\sim$50, where the model is still figuring out the atomic identities, and a ``relaxation'' regime, from diffusion step $\sim$50 to 0, where the model is moving around the atoms while holding the atomic species fixed.
The transition in ``Atom Elem. Final'' from 0\% finalized to 100\% finalized is fairly abrupt, suggesting that the model decides on all the atomic species at once, instead of first deciding on, e.g., the carbon structure and then deliberating about which functional groups to add.
Note also that the model finalizes geometries decidedly after choosing the atomic species: the curve showing which fraction of molecules have every atom's valence finalized (``Mol. BO Final'') doesn't increase at all until after almost all molecules' atomic species have been decided on (``Mol. Elem Final''), and the RMSD of interatomic distances rapidly decreases starting at diffusion step 50, right after the atomic species have mostly been decided upon. 

\begin{figure}
    \centering
    \includegraphics[width=\textwidth]{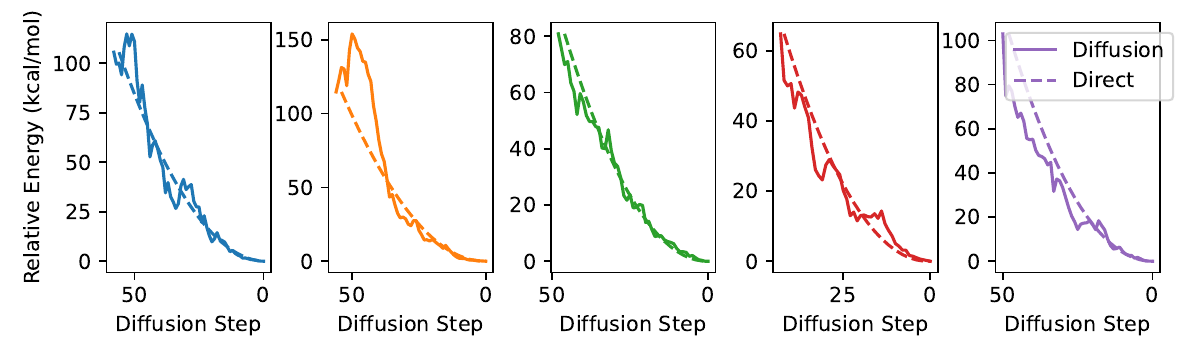}
    \caption{\textbf{Diffusion chain energies.} DFT-computed relative energy compared to the final generated geometry for the structures at the end of the diffusion chain that have the same atom types as the final structure. Each plot corresponds to a different generated molecule. Solid lines represent geometries predicted by diffusion, and dashed lines are geometries that are linearly interpolated between the initial and final geometries.}
    \label{fig:energy_vs_diffusion}
\end{figure}

Looking more closely at the relaxation regime, we calculate the energy of each structure along the diffusion path using DFT, starting with the first structure where all atom types are finalized.
As seen in Figure \ref{fig:energy_vs_diffusion}, the energy decreases during the relaxation regime fairly consistently, suggesting that the diffusion process is largely following the potential energy surface to the ground state rather than moving atoms through each other or taking a more erratic path.
For reference, the DFT-computed energies of the geometries along a linearly interpolated path from the starting structure to the final structure are shown by the dashed lines in Figure \ref{fig:energy_vs_diffusion}.
We investigate this further in \S \ref{sec:forces}.

\subsection{EDM Can Relax Structures}
\label{sec:relaxations}
Noting the separation of the diffusion process into these two regimes (``exploration'' and ``relaxation''), and that the energy of generated structures consistently decreases during the relaxation regime, a natural question to ask is: can we successfully relax arbitrary structures by running diffusion steps $\sim$50 $\to$ 0, starting at a user-chosen unrelaxed structure?
During training, the model only sees structures that are (roughly speaking) a Gaussian perturbation away from the ground state.
Any initial structure we provide the model that is not generated in this way may be outside of the training distribution.
As such, there's no guarantee that the relaxation regime will actually find a low-energy geometry if it is initialized at one of these out-of-distribution states.

\paragraph{Methods.}
To test this, we initialize the diffusion process with the ground state produced by running a structure optimization using the Merck Molecular Force Field (MMFF94) \cite{tosco2014bringing} on molecules from the QM9 validation set, and then we run the last $n$ steps of diffusion.
To find the MMFF ground state, we use RDKit \cite{landrum2023rdkit} to generate an initial guess for the 3D structure of the molecule given only the molecular graph, and then we relax this structure using MMFF.

The MMFF-optimized structures are already reasonable approximations of the DFT ground state,
but we seek to further improve them using EDM.
To evaluate the diffusion-generated structures, we compare the DFT-computed energies of the diffused structures with that of the MMFF structures.
Figure \ref{fig:qm9_samples} in Appendix \ref{app:sample_mols} shows a random sample of the molecules used in this section.

The question remains what diffusion step $n$ to start at when carrying out the relaxation.
As shown in Figure \ref{fig:diffusion_progression}, the model consistently makes smaller/larger steps for earlier/later diffusion steps, so we need to choose $n$ carefully: too small, and the model won't have enough steps to move the distance required to reach the ground state; too large, and the model will drastically re-arrange the molecule instead of finding the nearest local minimum.
Most likely, we should use $n<\sim 50$, since before this point, the model has not finalized atom species.
As such, we try three values of $n$: 50, 30, and 20.

\begin{figure}
    \centering
    \includegraphics[width=0.49\textwidth]{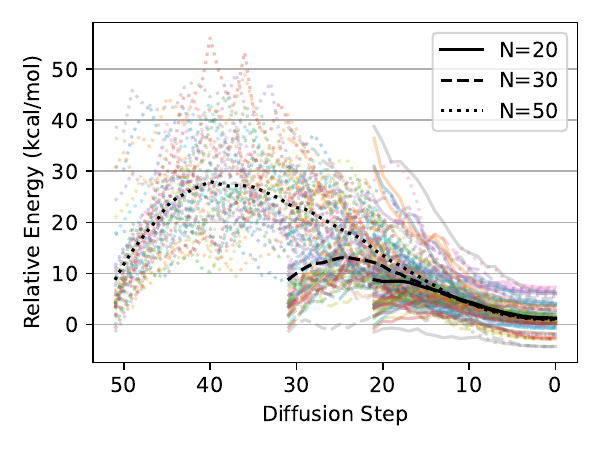}%
    \includegraphics[width=0.49\textwidth]{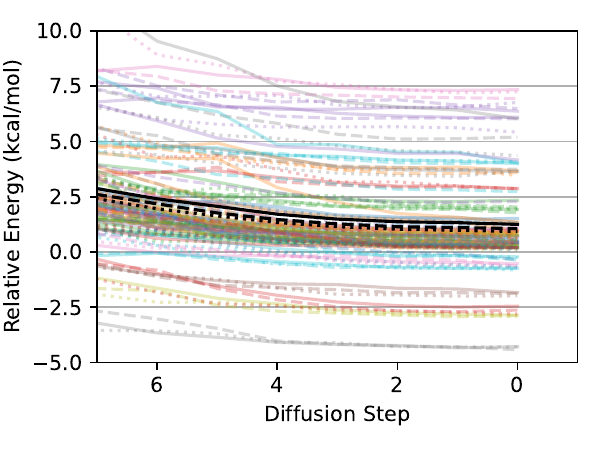}
    \caption{\textbf{Energies of EDM structural ``relaxations''.} Left: DFT-computed energies relative to the ground state for all structures on the diffusion-generated ``relaxation'' paths, for 45 randomly chosen molecules from the QM9 validation set. Each color is a different molecule from the QM9 validation set. Linestyle indicates how many diffusion steps were used ($n=20$, 30, and 50). Black lines are averages over all molecules. Right: Zoom of left figure. 
    }
    \label{fig:diffusion_relaxation}
\end{figure}

\paragraph{Results.}
Figure \ref{fig:diffusion_relaxation} shows DFT-computed relative energies between the DFT-relaxed structure and all structures along the diffusion path for each of these values of $n$.
The diffused structures are on average equally good for all values of $n$, on average about 1 kcal/mol worse than the DFT-computed ground state.
In contrast, the MMFF-optimized structures are nearly 10 kcal/mol higher in energy than the DFT-optimized structure.
On the low end, at $n=20$, the model immediately begins improving the structures, whereas on the high end, at $n=50$, the model first worsens the structures, increasing the energy, but eventually finds as good or better final structures as the lower values of $n$.
This is the same kind of behavior that we would expect from a more traditional optimization process, such as gradient descent with a learning rate inversely proportional to the optimization timestep.
As long as the initial learning rate is within a reasonable range, gradient descent will always converge to the local minimum, but a higher initial learning rate may cause the optimizer to initially worsen the objective.
Because in practice we do not know how far away our initial state is from the true ground state, it is important that this method is robust to choosing too many steps, since if so, we can safely overestimate $n$ and still arrive at high-quality ground states.

\subsection{Alignment of Denoising Steps with Forces}
\label{sec:forces}
As seen in Figure \ref{fig:energy_vs_diffusion}, the relaxation regime of the diffusion process consistently improves structures down to the ground state.
However, although the energy mostly decreases as diffusion proceeds, and the final structure is close to the DFT-computed ground state, it is unclear which relaxation path the model takes to the ground state.
One simple hypothesis is that it follows the forces down to the minimum; another is that it takes the most direct path to the minimum, compensating for the curvature of the PES along the way.
It is also possible that the model's predictions align neither with the forces nor with the direct path to the ground state, instead taking an erratic or circuitous path to the ground state.

\paragraph{Methods.}
To differentiate between these hypotheses, we compute the cosine similarity between the steps that the EDM model makes (``$\Delta$''), the DFT-computed forces at each geometry along the diffusion path (``$f$''), and the direction straight from each geometry to the DFT-relaxed ground state (``$gs$'').
If the model tends to follow the forces down to the minimum, then we expect the angle between $\Delta$ and $f$, (``$\theta_{\Delta, f}$'') to be small, or $\cos\theta_{\Delta, f}$ to be large.
In particular, it should be large compared to $\cos\theta_{\Delta, gs}$.
Similarly, if the model heads straight for the ground state, ignoring bumps in the PES along the way, then we expect $\cos\theta_{\Delta,gs}$ to be large compared to $\cos\theta_{\Delta,f}$.
We plot a third quantity, $\cos\theta_{f,gs}$, as a point of reference, since the forces tend to be somewhat aligned with the path directly to the ground state, especially for the simple PESs of QM9 molecules.
If $\Delta$ is less aligned both with $f$ and with $gs$ than $f$ is aligned with $gs$, then we may conclude that the model's predictions are not particularly aligned with either the DFT-computed forces or with the path directly to the ground state.

\begin{figure}
    \centering
    \includegraphics[width=0.4\textwidth]{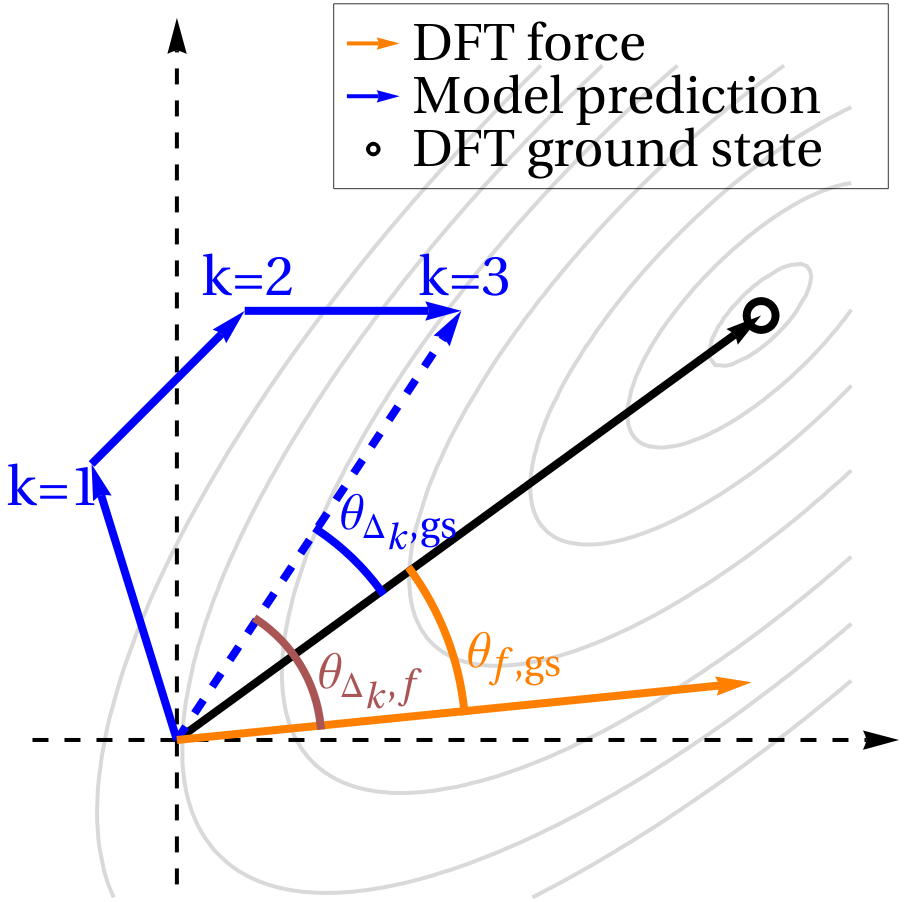}
    \caption{\textbf{Diagram explaining force comparisons.} Schematic showing what different values of $k$ mean.
    The origin of the plot is the current geometry under consideration, and the grey contours signify the DFT PES.
    $\Delta_k$ is the sum of the next $k$ diffusion steps taken by the model; $f$ is the DFT-computed forces on the atoms in the current geometry; $gs$ is the vector pointing from the current geometry to the ground state. 
    }
    \label{fig:k_schematic}
\end{figure}

However, noting that the diffusion process is inherently noisy, we also plot results when we consider the model prediction to be the sum of the subsequent $k$ steps of diffusion.
In other words, given $k\geq1$, $f$ and $gs$ remain unchanged, but instead of comparing these quantities to $\Delta$, we compare to $\Delta_k$, which is the sum of the next $k$ $\Delta$ vectors.
This idea is depicted schematically in Figure \ref{fig:k_schematic}.
For the smallest value of $k=1$, we expect the randomness in the diffusion steps to be most significant, so cosine similarities will likely be low.
Higher values of $k$ average out this noise, but raising $k$ also artificially increases $\cos\theta_{\Delta_k, gs}$, as compared to $\cos\theta_{\Delta_k, f}$, since if the model does eventually find a structure near the DFT ground state, regardless of which path it takes, $\Delta_k$ approaches $gs$ as $k$ increases.
For this reason, we plot three values of $k$: $k=1$, where the noise dominates, $k=30$, where $k$ is likely high enough that much of the similarity between $\Delta_k$ and $gs$ can be attributed to this effect, and $k=10$, which we hope strikes a good balance.

\begin{figure}
    \centering
    \includegraphics[width=0.49\textwidth]{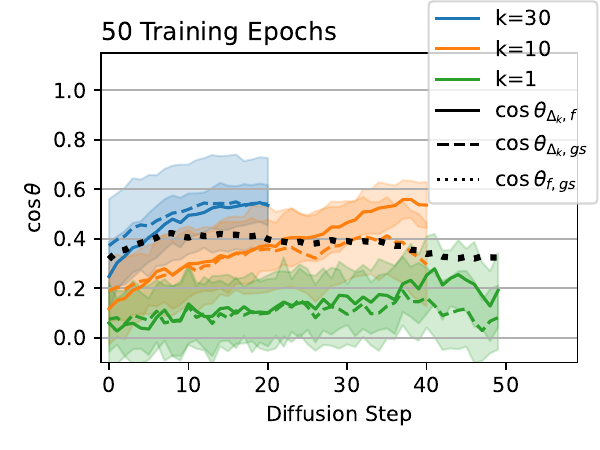}
    \includegraphics[width=0.49\textwidth]{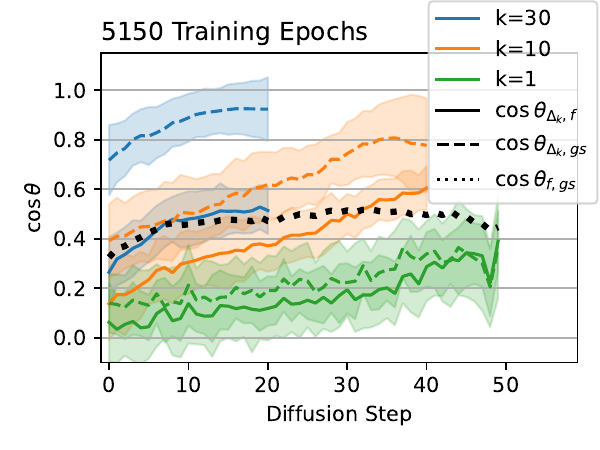}
    \caption{\textbf{QM9 Model/Force Comparison.} Left: cosine of the angles between the sum of the next $k$ steps predicted by diffusion ($\Delta_k$), the DFT-computed forces ($f$), and the path directly from the current geometry to the DFT-computed ground state ($gs$).
    Curves are averaged across atoms from 45 molecules from the QM9 validation set.
    Shaded regions represent one standard deviation above and below the mean.
    (The shaded region can exceed 1.0 because the distribution is not Gaussian.)
    }
    \label{fig:forces}
\end{figure}

\paragraph{Results.}
The results for the cosine similarity analysis are plotted in Figure \ref{fig:forces}.
The left plot shows results after only $50$ epochs of training, and the right plot shows results after the full $5150$ epochs.
As expected, for $k=1$, the green lines show $\cos\theta_{\Delta,f}$ and $\cos\theta_{\Delta,gs}$ are both fairly low throughout the inference process, due to the inherent randomness of diffusion.
However, as $k$ increases, alignment increases between the model prediction and both the forces and the direct path to the ground state.

Considering the fully trained model (5150 epochs), the model predictions $\Delta$ are consistently more aligned with the direct path to the ground state than with the DFT-predicted forces.
This holds true even for low $k$, suggesting that the model is finding a path to the ground state that is more efficient than following the forces gradient-descent style.
There is reasonably high alignment between the model predictions and the forces, but this can be explained by the fact that the forces themselves are somewhat aligned with the direct path to the ground state (black dotted line in Figure \ref{fig:forces}).
$\cos\theta_{\Delta_k,f}$ is never substantially higher than $\cos\theta_{f, gs}$, so any alignment between the model predictions and the forces can be explained by the alignment between the forces and the straight path to the ground state.
Given that these results are on molecules that were unseen during training, the fact that the diffusion path aligns better with the direct path to the ground state than with the forces suggests that the model has learned about local curvature of the PES, rather than only learning about local gradients.
Note that many non-learning-based relaxation methods also compensate for the curvature of the PES instead of strictly following the gradients down to a local minimum. 
For example, BFGS preconditions the gradients with second-order information in order to take a straighter path towards the minimum than would be achieved by following the gradients directly.

Surprisingly, the situation is nearly the reverse for the model that has only been trained for $50$ epochs (left plot of Figure \ref{fig:forces}).
Here, particularly later on in the diffusion process, $\Delta_k$ aligns better with $f$ than with $gs$, suggesting that the model is more so following the forces down to the ground state, rather than heading directly towards the ground state.
When training any model, early on in training, the model first picks up on the easiest way to get the answer mostly right, and later on in training learns to recognize more nuanced aspects of the input to make more nuanced predictions.
In this case, early on in training, the model moves the atoms preferentially in the same direction as the forces experienced by the atoms, despite having never seen any energy or force data during training.
In other words, the model has discovered ``following the forces'' as the easiest way to find a low-energy geometry (at least to the extent that the solid lines are higher than the dashed lines in the left side of Figure \ref{fig:forces}).
In contrast, by the end of training, the model learns to compensate for the curvature of the PES and to move directly towards the ground state, but learning to do so takes significantly more epochs of training.

\subsection{EDM as Boltzmann Generator}
\label{sec:boltzmann}
Figure \ref{fig:diffusion_progression} shows that diffusion takes monotonically decreasing step sizes as $n$ decreases.
Because there is inherent stochasticity in the diffusion process, the model does not always move directly towards a local minimum of the PES
(this is shown experimentally by the bumpiness in Figure \ref{fig:energy_vs_diffusion}).
But given that the model has a preference for low-energy geometries, there is a potential for repeated application of the same diffusion step to a structure to result in sampling low-energy geometries more often than high-energy geometries.
In particular, we might hope that the distribution of geometries follows the Boltzmann distribution, and that different diffusion steps $n$ would correspond to particular temperatures $T$.

\paragraph{Methods.}
To investigate this possibility, we carry out Markov chain Monte Carlo (MCMC) simulations on ten molecules randomly from the QM9 validation set after filtering to select for flexible and linear molecules, where we expect more interesting behavior at non-zero temperature.\footnote{In particular, we use a simple filter based on the molecules' SMILES strings: we filter out SMILES containing ``='' or ``\#'', SMILES containing numbers, and SMILES with more than 15 ``('' (to bias towards more linear molecules).}
We run the Metropolis-Hastings algorithm for 24,000 steps with an isotropic Gaussian proposal distribution---initialized at the DFT ground state with 5,000 steps of burn-in---using eighteen temperatures $T$ between 10 K and 400 K.
On the diffusion side, we repeatedly apply a single step of diffusion to the structures 20,000 times, initialized at the DFT ground state with 1,000 steps of burn-in (we observed that the chain stabilizes significantly faster than for Metropolis-Hastings).
We try fourteen different steps $n$, from the $n=1$ up to $n=40$.
In both cases, we use GFN2-xTB \cite{bannwarth2019gfn2} to measure energies, as implemented in the \texttt{xtb-python} package (version 22.1).

For very low $T$ and $n$, we expect both the MCMC simulation at temperature $T$ and the diffusion chain with step $n$ to have an energy of $\sim0$ relative to the ground state.
However, because there is sometimes disagreement between GFN2-xTB and the DFT-calculated ground states that the model was trained on, the diffusion chains at low $n$ settle to an xTB-computed energy slightly above zero.
To compensate for this, for each molecule we subtract a constant energy ($<2$ kcal/mol) from each chain to equalize the minimum energies achieved by the two chains at the lowest values of $T$ and $n$.

\begin{figure}
    \centering
    \includegraphics[width=\textwidth]{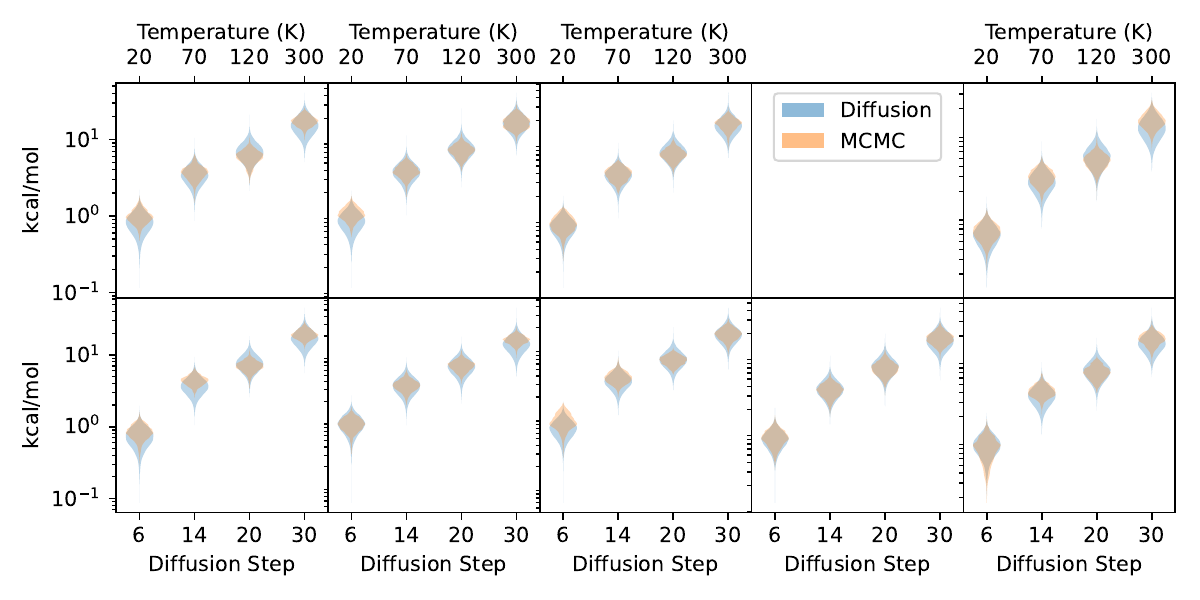}
    \caption{\textbf{Comparing MCMC and diffusion.} Violin plots showing the distribution of energies obtained with a Metropolis-Hastings MCMC simulation (orange), along with energies obtained by repeatedly applying a single diffusion step to a structure (blue). Temperatures used for the MCMC simulation are on the top axis, and diffusion steps used are on the bottom axis.}
    \label{fig:mcmc_hists}
\end{figure}

\paragraph{Results.}
Histograms of the chain energies for selected values of $T$ and $n$ are shown in Figure \ref{fig:mcmc_hists}.
These values of $T$ and $n$ were chosen to maximize the overlap between the distributions, but they are the same for each of the nine molecules,\footnote{We tested on ten molecules, but we discarded one because GFN2-xTB disagrees strongly with DFT about where the local minimum of the PES is.} suggesting that there may be a correspondence between $n$ and $T$ that generalizes across molecules.
Within each molceule, as $T$ and $n$ increase, we see the same trend of the mean relative energy increasing and the variance of the distribution increasing.

As a point of reference, when repeatedly perturbing atomic coordinates with an isotropic normal distribution instead of the diffusion model, the energy diverges, even for extremely small step sizes.
Unlike repeated Gaussian perturbations, which have no preferred direction, the diffusion model preferentially moves the geometries closer to a local minimum of the PES, even without the Metropolis-Hastings acceptance criterion, which uses a ground-truth energy oracle.
The model was trained only on geometries, with no energy supervision either on the ground-state geometries or on any non-equilibrium geometries.
Despite this, we never observed a diffusion chain diverging, even for high values of $n$, and there is even reasonable agreement between the distributions of energies within the MCMC and diffusion chains.

\begin{figure}
    \centering
    \includegraphics[width=0.49\textwidth]{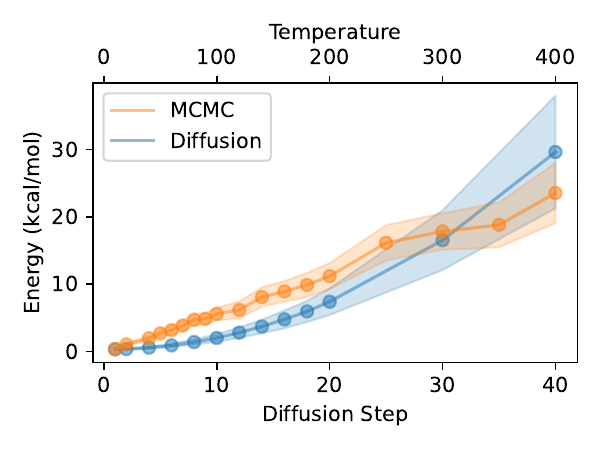}%
    \includegraphics[width=0.49\textwidth]{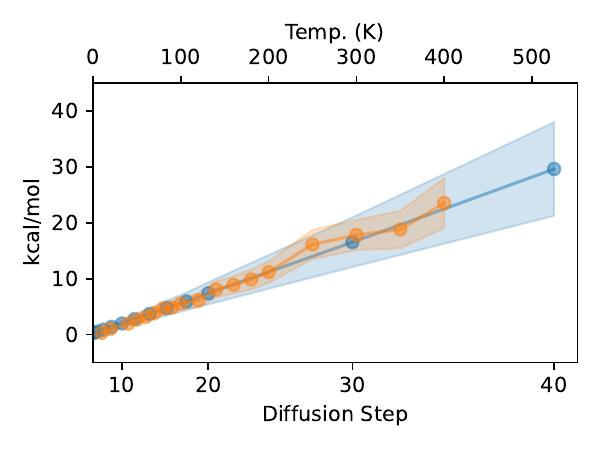}
    \caption{\textbf{Quadratic relationship between diffusion step and energy.} Left: average energy of the MCMC and diffusion chains, with shaded regions corresponding to one standard deviation above and below the mean.
    Values on the $x$-axis indicate the temperature used in the MCMC simulation (top) and the diffusion step $n$ used when making the diffusion chain.
    Right: Same as left, but the lower $x$-axis is scaled quadratically and stretched linearly to match the slope of the MCMC line
    }
    \label{fig:mcmc_mol0}
\end{figure}

\begin{figure}
    \centering
    \includegraphics[width=\textwidth]{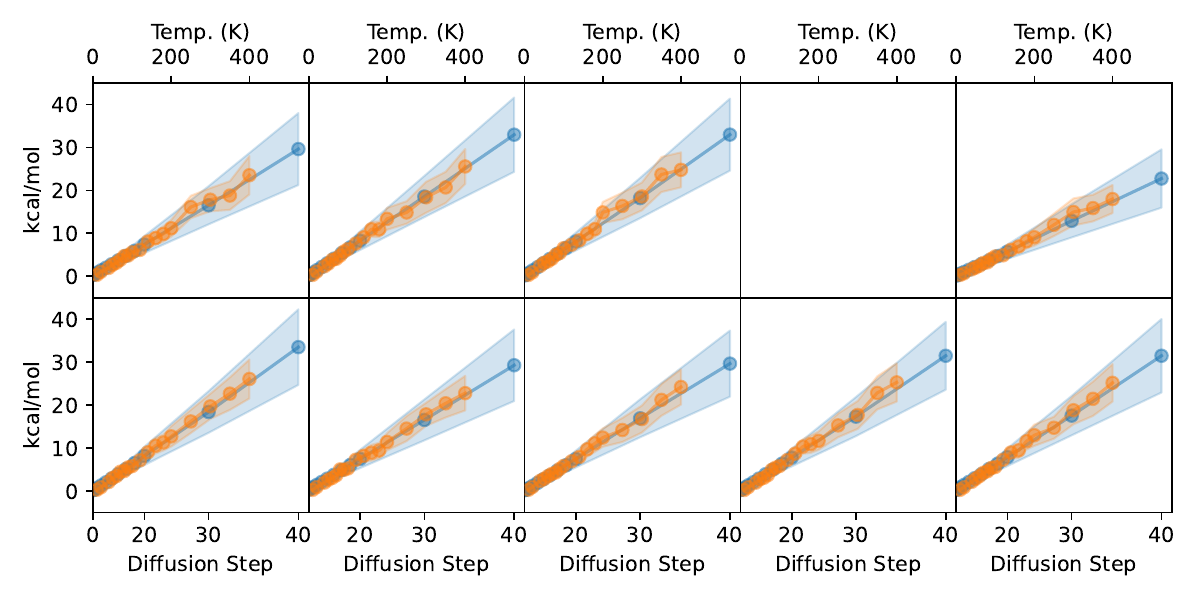}
    \caption{Right-hand plot of Figure \ref{fig:mcmc_mol0} for the nine molecules considered. Only one linear scaling factor is used across all molecules.}
    \label{fig:mcmc_sqrt}
\end{figure}

Next we investigate the relationship between $T$ and $n$ as well as the average and variance of the resulting energy distributions.
Figure \ref{fig:mcmc_mol0} plots, for a single molecule, the average energy $\pm$ the standard deviation of the energy for each of the $n$ and $T$ values we considered.
As expected, the energy increases linearly with temperature.
On the other hand, the energy increases quadratically with increasing $n$.
This is unsurprising: near the end of inference, the step size decreases roughly linearly, as seen in Figure \ref{fig:diffusion_progression}, and near a local minimum, we expect the PES to be modeled well as a harmonic oscillator.
The left side of Figure \ref{fig:mcmc_mol0} plots both chains with a linear scale on $T$ and $n$.
The right side of the figure instead uses a quadratic scale for $n$, and the $x$-axis is linearly scaled by a constant $\mu$ to equalize the slope between the two chains.
Note that any linear and quadratic functions can be made to line up using this method, so their alignment in this plot is unsurprising.
Figure \ref{fig:mcmc_sqrt} plots the same quantities, but repeated for each of the nine molecules considered.
In this case, we use the same linear scaling $\mu$ for each molecule, so there is no guarantee that the lines will all line up.
Even though there is some variation in the heat capacities of the nine molecules (i.e. the slope of the MCMC lines), the diffusion chain consistently generates very similar average energies as the MCMC chain at the corresponding temperature.

The variances of the distributions are also similar, though the diffusion chain consistently results in a wider distribution of energies than the MCMC chain.
The energy distributions for these simple molecules from QM9 are unimodal, but it would be interesting to see whether the diffusion chain can reproduce multi-modal energy distributions for more complex molecules.

\section{Using Equivariant Diffusion Models To Accelerate Relaxations}
\label{sec:accelerate}
\subsection{Diffusion-relaxed Structures Accelerate DFT Relaxations}
\label{sec:speedup}
This completes our investigation of what EDM models are learning about the PES.
Next, we turn to a practical application that, to our knowledge, has not been explored with this sort of self-supervised model: the acceleration of DFT structure relaxations by proposing better initial geometries than those obtained with a classical force field.
In \S \ref{sec:relaxations}, we show that the geometries predicted by EDM have significantly lower energy than those predicted by MMFF.
We may therefore hope that DFT relaxations starting at the EDM geometries will converge faster than relaxations starting at the MMFF-produced geometries.

\begin{figure}[!ht]
    \centering
    \includegraphics[width=0.52\textwidth]{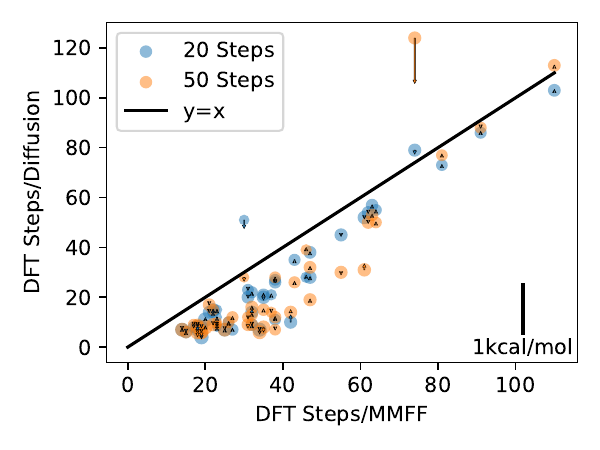}
    \includegraphics[width=0.47\textwidth]{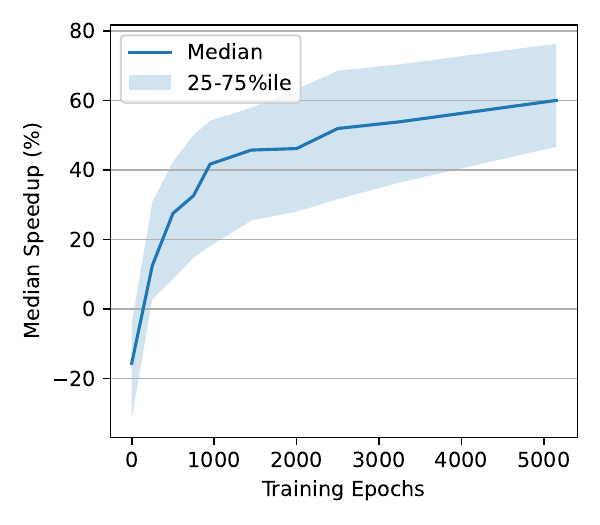}
    \caption{\textbf{QM9 DFT relaxation speedup.} Left: Each point is a structure that we relaxed using DFT.
    $x$ values are the number of relaxation steps required to converge when initializing DFT with the MMFF ground state.
    $y$ values are the number of steps when initializing with the structure predicted by diffusion.
    Orange/blue dots used $n=50$/$n=20$ steps of diffusion.
    Arrows indicate the difference between the energy of the DFT-relaxed structure when starting from the MMFF structure vs. starting from the diffused structure.
    Downward-pointing arrows indicate that the relaxed structure had lower energy when the DFT relaxation was initialized with diffusion rather than with MMFF.
    The area of each circle is proportional to the number of atoms in the structure.
    Right: median speedup attained when using checkpoints throughout training.
    The left plot shows results for epoch 5150.
    }
    \label{fig:speedup}
\end{figure}

\paragraph{Methods.}
To test this, we compare the number of DFT relaxation steps required to relax structures when initialized at the MMFF ground state vs. at the diffusion-generated geometry.
We perform DFT relaxations from three different starting points: the MMFF-optimized structure, and the structures obtained by further ``relaxing'' this MMFF-optimized configuration using diffusion with both $n=20$ and $n=50$.
We also compare the energies of the DFT-relaxed structures when starting at each of these starting configurations.

\begin{figure}
    \centering
    \includegraphics[width=0.8\textwidth]{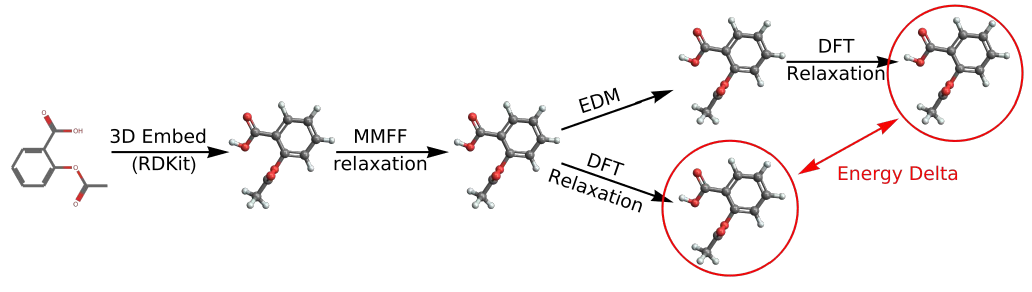}
    \caption{Schematic showing what DFT relaxations we carry out and how we calculate the energy delta.}
    \label{fig:relaxation_diagram}
\end{figure}

In addition to measuring the speedup obtained by starting at the diffusion-generated structure instead of the MMFF-relaxed structure, we also compare the energies of the final structures themselves after undergoing DFT relaxation.
In most cases, DFT finds nearly identical structures, regardless of which of the two initial structures were used.
However, in some cases, the two structures do differ, and we quantify this difference using the relative energy between the two structures.
In the following sections, we refer to this relative energy as the ``energy delta.''
An energy delta greater than zero indicates that the DFT relaxation converged to a higher-energy structure when initialized at the MMFF-relaxed structure than at the diffusion-generated structure.
Figure \ref{fig:relaxation_diagram} shows a schematic of the overall calculation workflow, with the structures used to calculate the energy delta circled in red. 

\paragraph{Results.}
The number of DFT-based geometry optimization steps required for $n=20$ and $n=50$ are shown on the left of Figure \ref{fig:speedup}, plotted vs. the number of steps required when starting at the MMFF-optimized structure.
For both $n=20$ and $n=50$, the DFT relaxations are consistently faster: the median number of steps required is 40\% lower for $n=20$ and 57\% lower for $n=50$.
On each point in Figure \ref{fig:speedup}, we plot an arrow that indicates the energy delta (positive values correspond to upward-pointing arrows).
If the diffusion-generated geometries stay within the same local minimum of the DFT potential energy surface as the MMFF-relaxed structures, then we expect the energy delta to be approximately zero (the scale bar for the arrows is shown in the lower right of the figure).
A downward-pointing arrow means that the structure obtained by relaxing the diffusion-generated structure using DFT resulted in a lower energy than that of the structure obtained by relaxing the MMFF-relaxed structure.
In other words, a downward-pointing arrow indicates that diffusion moved the geometry into a deeper basin of the DFT potential energy surface than the basin found by MMFF.

The energy deltas are approximately zero in every structure but one, in which diffusion with $n=50$ steps moved the geometry to a better local minimum than MMFF found, as indicated by the negative arrow of magnitude $\sim$1 kcal/mol. 
This structure is also one of the few where the relaxation takes longer when starting from the diffusion-generated geometry.
We visualize this particular structure in Appendix \ref{app:outlier}, where it becomes clear that diffusion has rotated the hydrogen atoms of a methyl group just far enough around the central carbon to get relaxed by DFT to a more favorable position.
Interestingly, when using only $n=20$ steps on the same structure, diffusion does not stray from the PES basin found by the MMFF relaxation, which is unsurprising given the results in Figure \ref{fig:diffusion_relaxation}.

We also calculate the median speedup obtained using EDM checkpoints throughout training.
Results are shown on the right of Figure \ref{fig:speedup}.
The improvement slows down over the course of training, but there may still be further improvements possible by simply training longer.
By the end of training, while in some cases the diffusion-generated structure leads to a slowdown rather than a speedup, even the 25$^{th}$ percentile speedup is nearly $2\times$.
For 25\% of structures, the speedup is at least $4\times$ (75$^{th}$ percentile is at 75\% speedup).

\subsection{Exploring Larger Molecules with GEOM}
\label{sec:geom}
Clearly, for the very small molecules in the QM9 dataset, diffusion-generated structures are generally significantly better initialization points for DFT relaxations.
Now, we turn to the larger and more realistic structures found in the Geometric Ensemble Of Molecules (GEOM) dataset \cite{axelrod2022geom}.
We use the ``drugs'' subset of GEOM, which consists of drug-like molecules, most of which are significantly larger than the molecules in QM9.
The typical number of heavy atoms in a conformation drawn from the drugs subset of the GEOM dataset is $22-29$ (25$^{th}$-75$^{th}$ percentile), whereas QM9 contains structures with only up to 9 heavy atoms. 
In this manuscript, we refer to the drugs subset of the GEOM dataset as simply the \drugs{} dataset.

\paragraph{Methods.}
We carry out the same speedup analysis as in \S \ref{sec:speedup} but using structures from the \drugs{} dataset.
Instead of calculating energies and performing relaxations using DFT, we use GFN2-xTB, both to match how the \drugs{} dataset was created and to save on computational cost.
As above, all relaxations are carried out with \texttt{fmax} $=0.03$.
We train an EDM model on the \drugs{} dataset using the same hyperparameters as used by \textcite{hoogeboom2022equivariant}.
However, \textcite{hoogeboom2022equivariant} train for only 13 epochs, or 1.2 million iterations, whereas we train for an additional 250 epochs, or about 23 million iterations, starting at their pretrained model.
Training takes 2.8 hours per epoch on 8 NVIDIA Quadro RTX 6000 GPUs.
The validation loss continues to improve throughout training, so additional training time is likely to increase the quality of the model.
Figure \ref{fig:geom_samples} in Appendix \ref{app:sample_mols} shows a random sample of the molecules used in this section.

\begin{figure}
    \centering
    \includegraphics[width=0.52\textwidth]{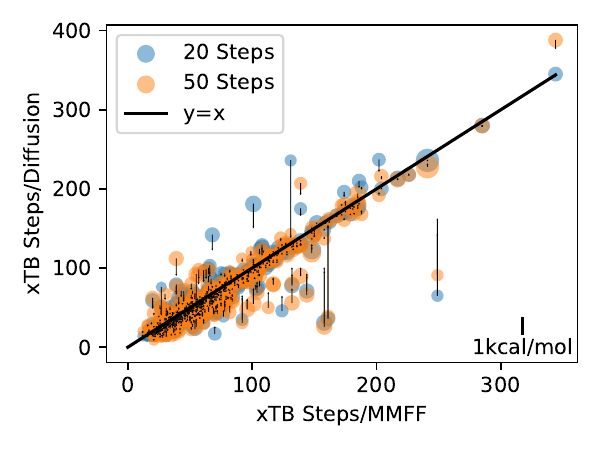}
    \includegraphics[width=0.47\textwidth]{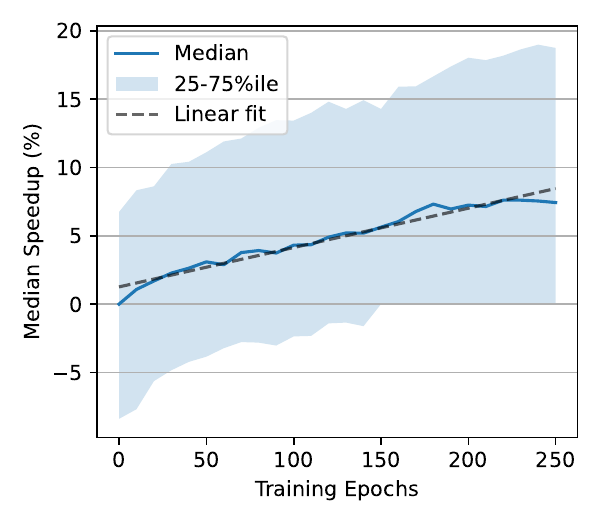}
    \caption{Speedup results for \drugs{}. Figures are analogous to those in Figure \ref{fig:speedup}.}
    \label{fig:speedup_geom}
    
\end{figure}

\paragraph{Results.}
As when trained on QM9, EDM finds structures with lower energies than those found by relaxing with MMFF.
However, the reduction in energy is smaller than for QM9: the diffusion-generated structures have an average energy relative to the xTB ground state of 8.3 kcal/mol vs.\ MMFF's 14.3 kcal/mol---a 42\% reduction, compared to nearly $90\%$ for QM9.
The median speedup to xTB relaxations by starting at these diffusion-generated structures instead of the MMFF structures is correspondingly smaller, at 7\%, compared to a 57\% speedup to DFT relaxations for QM9.
Figure \ref{fig:speedup_geom} shows results for \drugs{} analogous to Figure \ref{fig:speedup}.
The speedup attained increases nearly linearly with the number of training epochs, but it does appear to start tapering off by epoch 250.
As with the model trained on QM9, structures generated by diffusion sometimes cause the DFT relaxation to find a different local minimum of the PES; this can be seen by the points with high-magnitude arrows in the left side of Figure \ref{fig:speedup_geom}.
However, the speedup we observe is not due to the model finding worse local minima that take fewer steps to reach the bottom: when excluding any structure that led to an absolute energy delta $>0.2$ kcal/mol, the right side of Figure \ref{fig:speedup_geom} remains nearly unchanged except a slight reduction in variance (see Appendix \ref{app:lowE}).
Although further training is likely to lead to some improvements, most likely, improvements in model architecture or training hyperparameters would be needed to attain significantly higher speedups.

\begin{figure}
    \centering
    \includegraphics[width=0.49\textwidth]{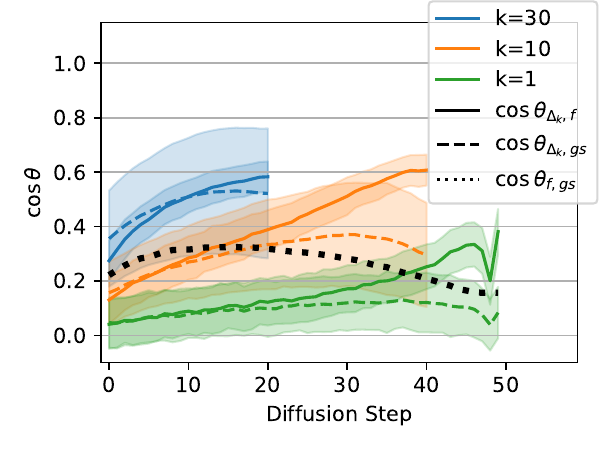}
    \caption{Force alignment plot, analogous to the plots in Figure \ref{fig:forces} but for the EDM model trained on \drugs{} for 250 epochs, and using xTB to calculate forces instead of DFT.}
    \label{fig:forces_geom}
\end{figure}

Lastly, we carry out a similar analysis of how the steps taken by diffusion align with the ground-truth forces on the atoms, using xTB to calculate forces instead of DFT.
For QM9 structures, early in training the model's predictions align better with the DFT-computed forces, and later in training they align better with the direct path to the ground state (\S \ref{sec:forces}).
In the case of \drugs{}, the model behaves similarly to how the QM9 model behaves early in training; the model's predictions are more aligned with the xTB-computed forces than with the path directly to the ground state.
This is shown Figure \ref{fig:forces_geom}, where the solid lines tend to go higher than the dashed lines.

This result helps paint a picture of how these models improve throughout training.
Early on, the QM9 model's predictions align better with the forces than with the direction to the ground state, and the structures produced by the model are only slightly better than those produced by MMFF:
after 50 epochs of training, the QM9 model reduces energy relative to the DFT ground state by $2\times$ compared to MMFF, and using the diffusion-produced structure yields no speedup to DFT relaxations whatsoever (after 100 epochs, the energy reduction is $3\times$, the median DFT speedup is 4\%, and the force alignment plot looks similar to the plot for 50 epochs).
Later on in training, the model's predictions trend towards pointing straight to the ground state instead of aligning with the forces, the energy improvement is close to $10\times$, and the median speedup to DFT relaxations is nearly 60\%. 
The \drugs{} results mimic the QM9 results early on in training: the model predictions align better with the forces than with the path to the ground state, the energy relative to the xTB ground state is reduced by a factor of $\sim 2$ compared to MMFF, and there is only a few percent speedup to xTB relaxations.
With a better model---whether through further training of this same EDM or after improving the model architecture---we might expect the model to behave more like the QM9 model behaves at the end of training.

\section{Conclusions}

We investigate how much can be learned about the PES by training a diffusion model on only ground state geometries. Our investigation is motivated by the need to lessen our reliance on supervised datasets generated via physics-based calculations. 
The trained model is able to relax QM9 structures to 10x lower relative energy than MMFF, and it can produce structures that require 57\% fewer DFT steps to relax than MMFF-relaxed geometries.
When ``relaxing'' QM9 structures, the model follows a noisy estimate of the path directly to the ground state rather than taking the path of steepest descent, suggesting that it has learned about at least the second-order curvature of the PES near the local minimum.
The model can also sample geometries around a local minimum in the PES from a distribution that resembles the Boltzmann distribution, and it can model the varying heat capacity of different materials.

In contrast, for more complex and larger GEOM-DRUGS molecules, the model's predictions align better with the ground-truth forces on the atoms, suggesting the model has not yet learned beyond first-order information about the PES.
Correspondingly, geometries predicted by EDM are of lower quality, both in terms of the energy improvement over MMFF and in terms of the speedup in DFT relaxations when starting at the EDM-predicted structures.

Although we present some of our findings in terms of capabilities that the model has, we are not proposing that a self-supervised model could outperform state-of-the-art supervised NNIPs on tasks like structure relaxations or MCMC simulations.
After all, the training data for the diffusion model used in this manuscript consists of only a single point on the PES for each molecule or conformation. 
NNIP training datasets, in contrast, contain many points on the PES for each molecule, all labeled with energies and forces.
Rather, we investigate the model's capabilities as a way both to see how far it is possible to get with self-supervision alone, and to gain insight into what information these models are learning about the PES from a training set of only ground-state geometries.

We see a number of avenues for future work.
One exciting direction is to explore new tasks that have traditionally required an interatomic potential but that could be carried out with a self-supervised model instead.
For example, in Section \ref{sec:boltzmann}, we establish a correspondence between diffusion steps and temperature; future work could make use of this correspondence to use EDM for replica exchange MCMC.
In a similar vein, repeatedly applying EDM to a structure while progressively increasing the diffusion step could allow the model to accelerate reaction prediction calculations and/or transition state estimation.
Future work could also explore whether our findings generalize to different types of materials, such as crystals and glasses.
Lastly, on the modeling side, future work could explore other existing denoising models, and could also develop new model architectures and training paradigms designed specifically to improve performance on, e.g., structure relaxations.

\paragraph{Acknowledgments.}
A. S. R. acknowledges support via a Miller Research Fellowship from the Miller Institute for Basic Research in Science, University of California, Berkeley.
A. S. K. acknowledges support from Laboratory Directed Research and Development (LDRD) funding under Contract Number DE-AC02-05CH11231. 
D. R. and J. E. G. acknowledge gifts from Anyscale, Astronomer, Google, IBM, Intel, Lacework, Microsoft, Mohamed Bin Zayed University of Artificial Intelligence, Samsung SDS, Uber, and VMware.

\newpage
\setlength{\emergencystretch}{25pt}
\printbibliography
\setlength{\emergencystretch}{0pt}

\newpage

\appendix

\section{Sample Molecules}
\label{app:sample_mols}
We visualize sample molecules that were used in the analyses from the main text.
All structures shown are the geometries from the respective datasets (QM9 and \drugs{}).
Colors are as follows: Hydrogen (white), Carbon (grey), Oxygen (red), Nitrogen (blue), Chlorine (light green), Fluorine (dark green), Sulfur (yellow), Bromine (brown).

\begin{figure}[h]
    \centering
    \begin{tabular}{ccccc}
    \includegraphics[width=0.09\textwidth]{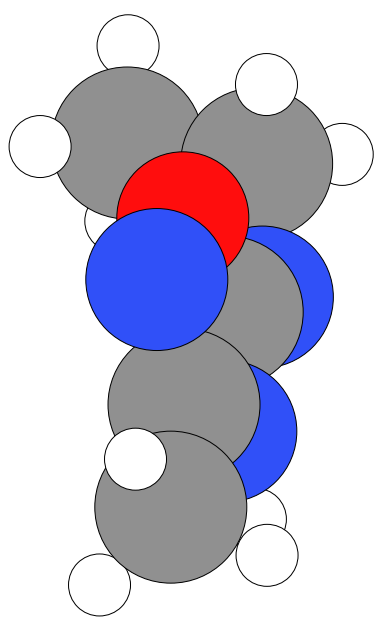} &
    \includegraphics[width=0.11\textwidth]{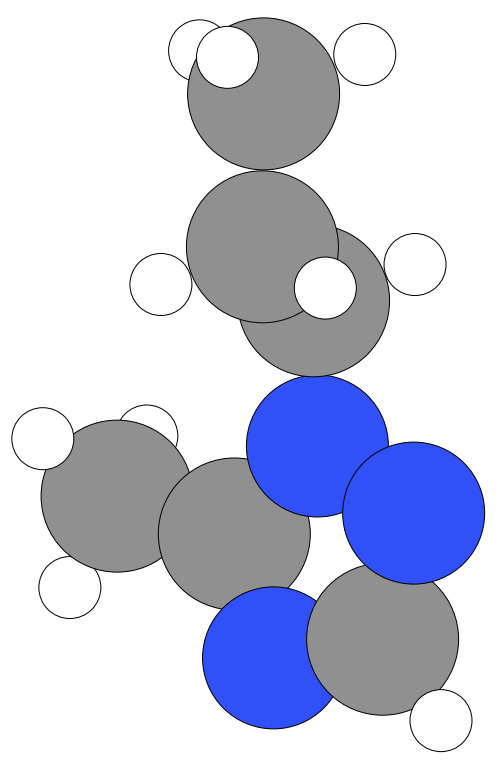} &
    \includegraphics[width=0.11\textwidth]{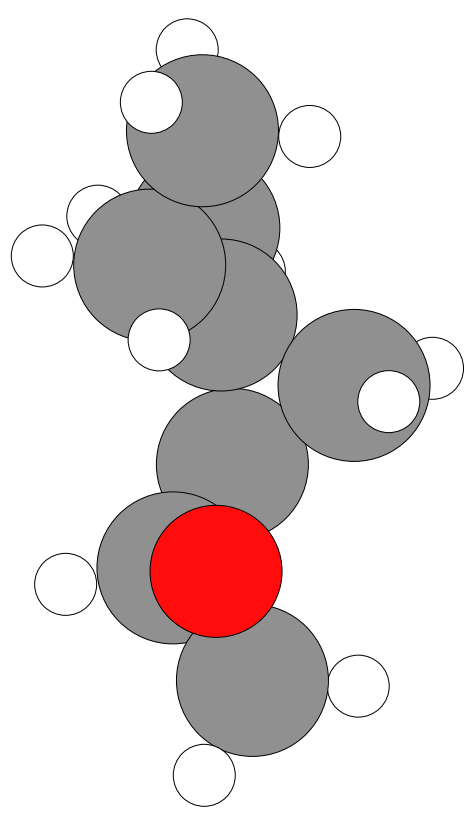} &
    \includegraphics[width=0.14\textwidth]{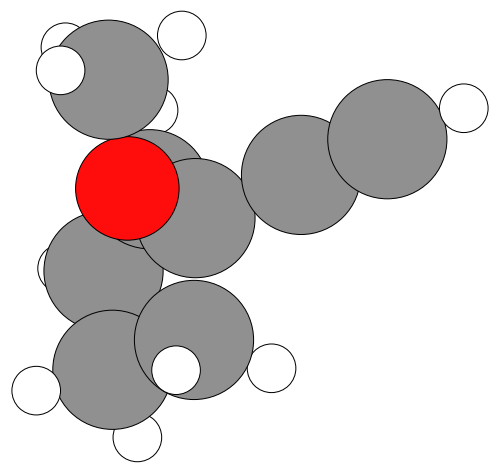} &
    \includegraphics[width=0.15\textwidth]{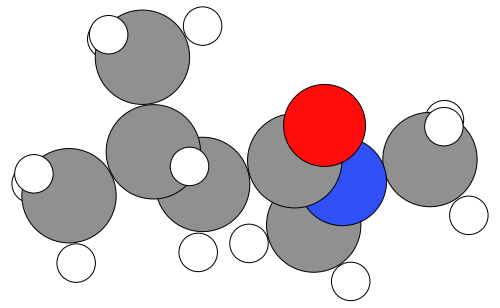}
    \\
    
    \includegraphics[width=0.12\textwidth]{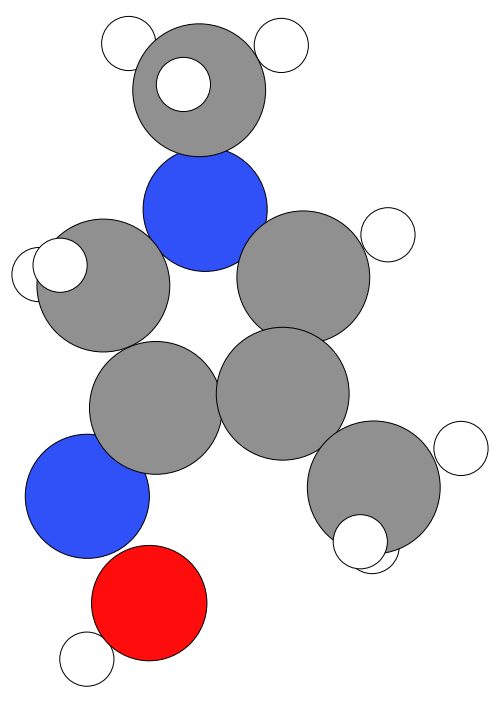} &
    \includegraphics[width=0.12\textwidth]{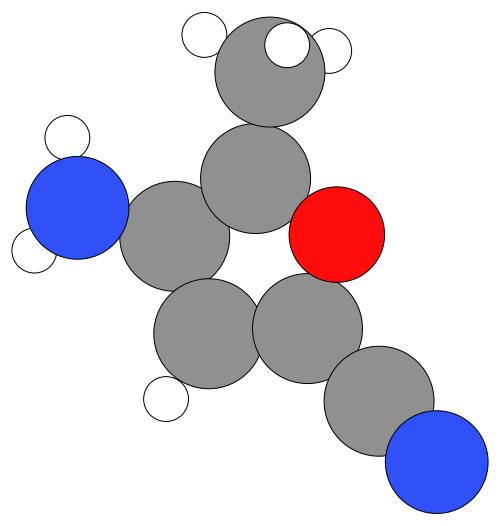} &
    \includegraphics[width=0.10\textwidth]{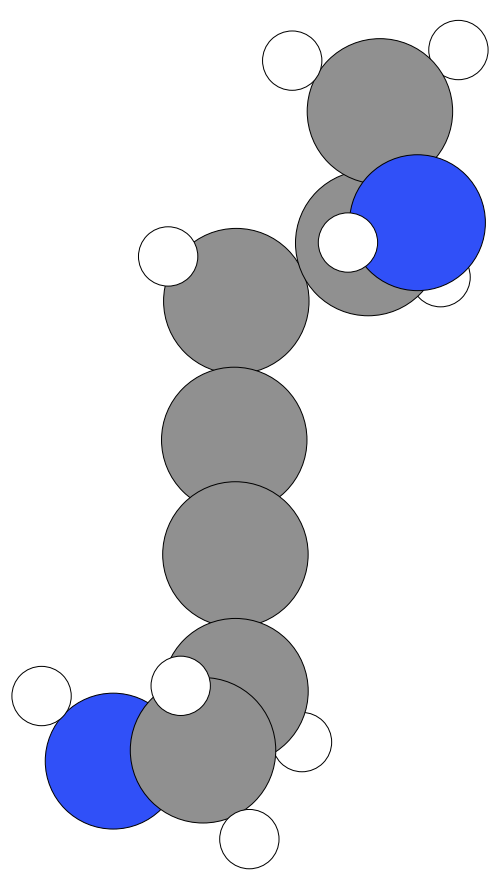} &
    \includegraphics[width=0.14\textwidth]{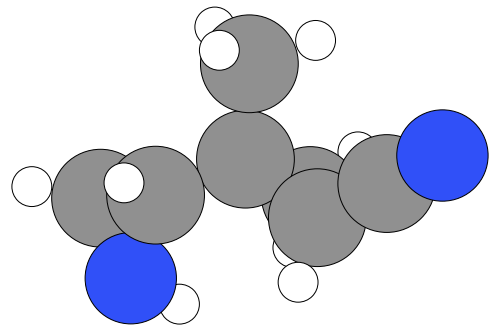} &
    \includegraphics[width=0.12\textwidth]{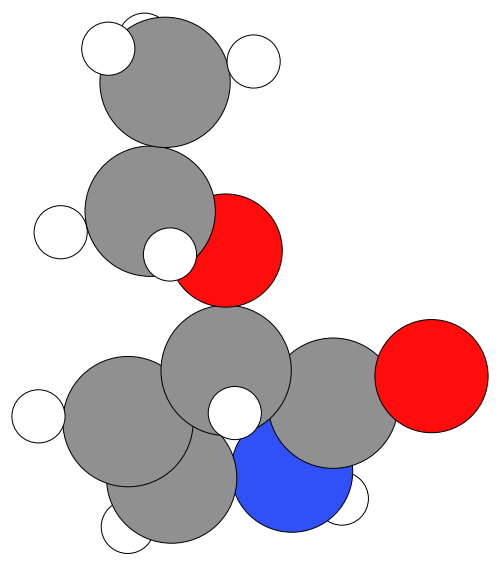}
    \\

    \includegraphics[width=0.12\textwidth]{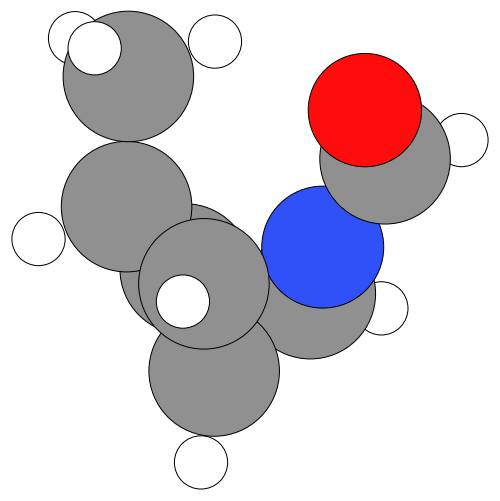} &
    \includegraphics[width=0.12\textwidth]{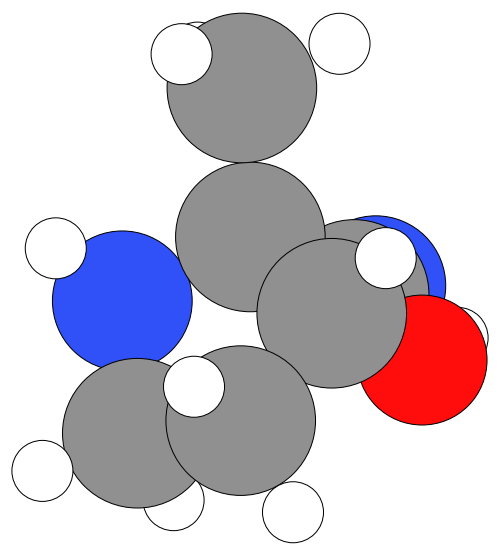} &
    \includegraphics[width=0.12\textwidth]{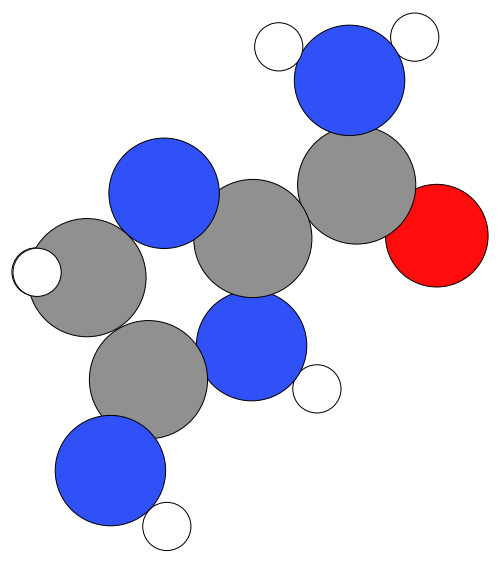} &
    \includegraphics[width=0.12\textwidth]{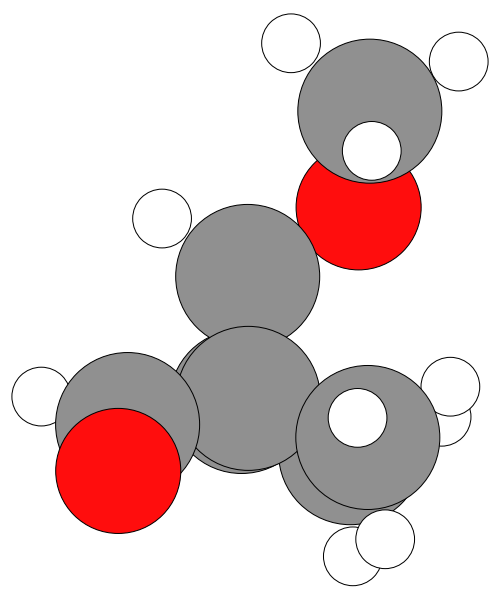} &
    \includegraphics[width=0.12\textwidth]{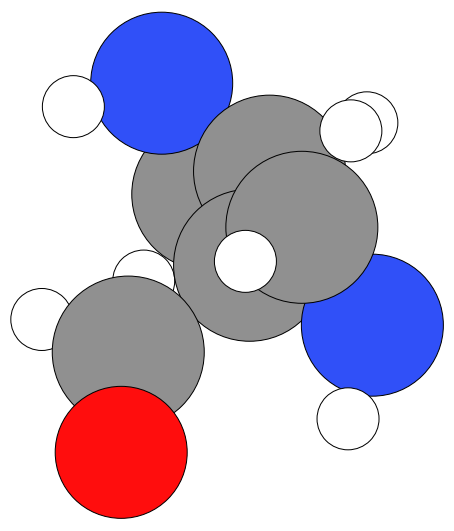}
    \\
    
    \includegraphics[width=0.12\textwidth]{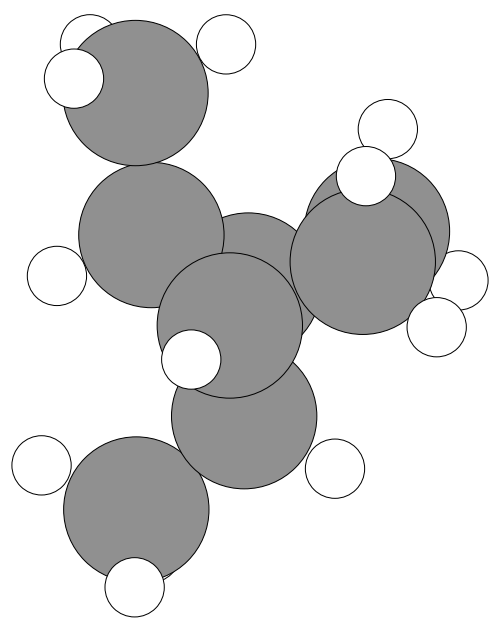} &
    \includegraphics[width=0.12\textwidth]{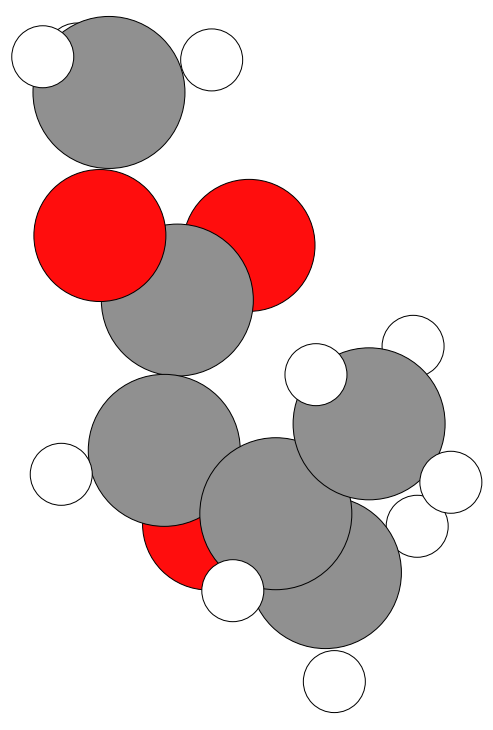} &
    \includegraphics[width=0.12\textwidth]{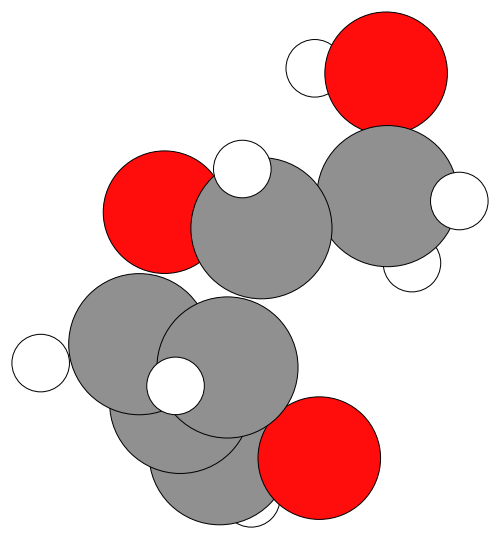} &
    \includegraphics[width=0.12\textwidth]{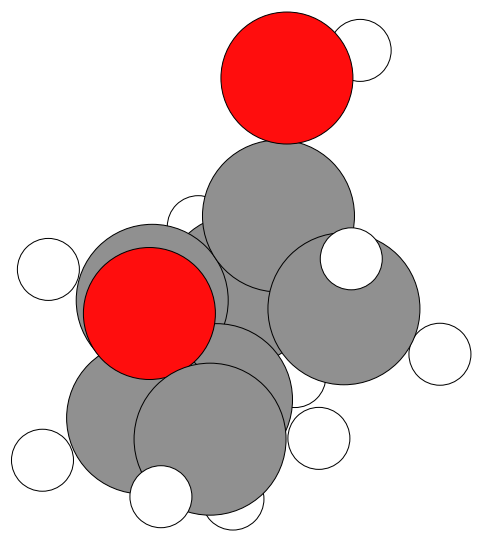} &
    \includegraphics[width=0.12\textwidth]{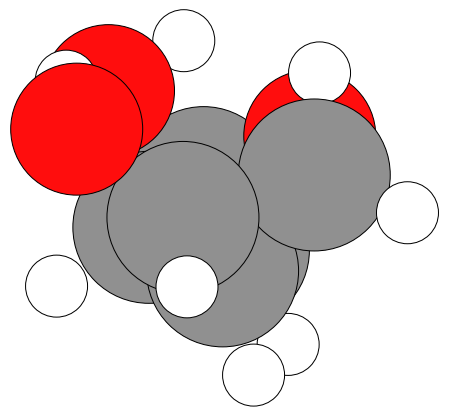}
    \\
 
    \end{tabular}
    \caption{Samples from QM9}
    \label{fig:qm9_samples}
\end{figure}

\begin{figure}[H]
    \centering
    \begin{tabular}{cccc}
    \includegraphics[width=0.21\textwidth]{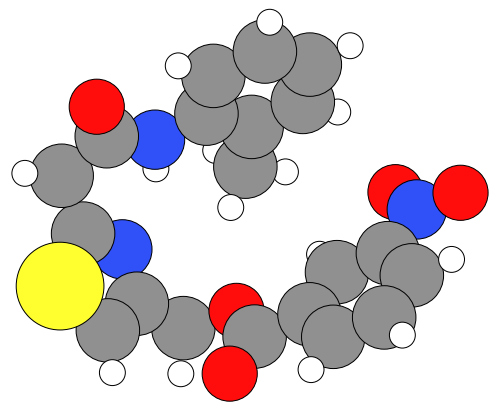} &
    \includegraphics[width=0.21\textwidth]{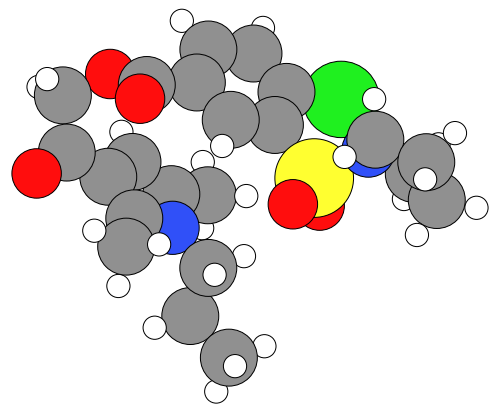} &
    \includegraphics[width=0.21\textwidth]{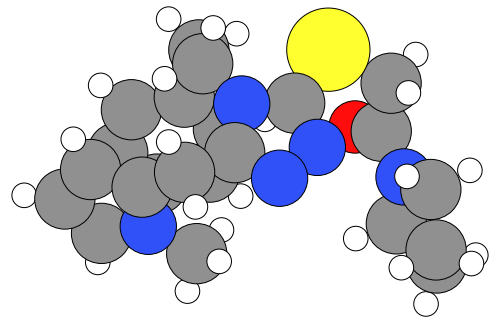} &
    \includegraphics[width=0.21\textwidth]{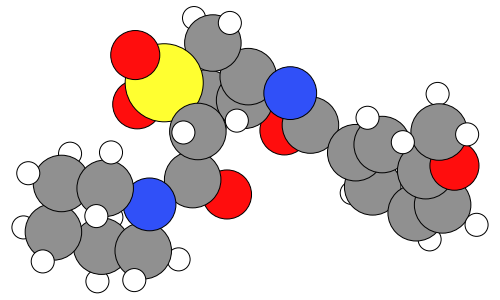}
    \\

    \includegraphics[width=0.21\textwidth]{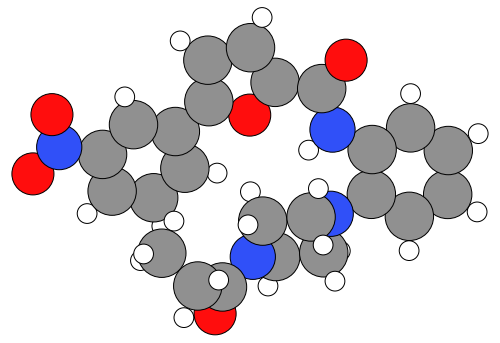} &
    \includegraphics[width=0.21\textwidth]{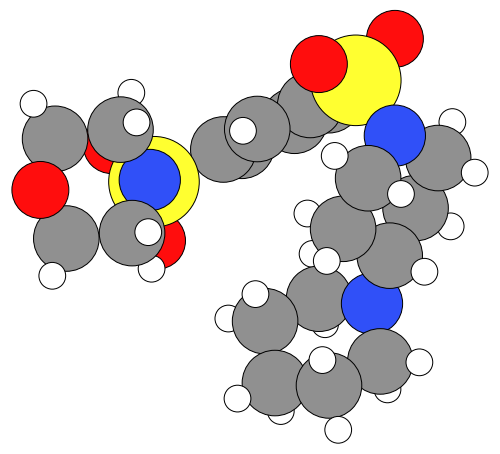} &
    \includegraphics[width=0.21\textwidth]{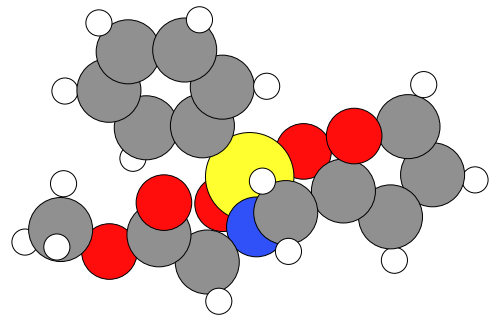} &
    \includegraphics[width=0.21\textwidth]{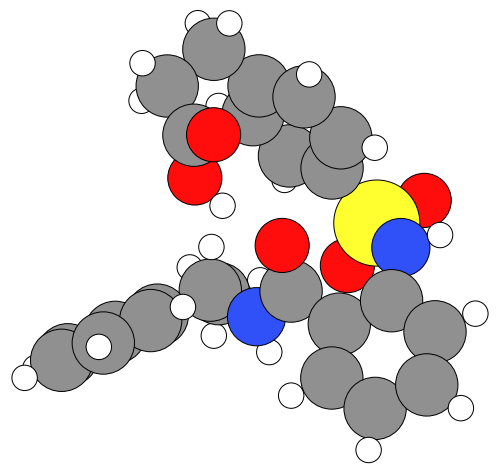}
    \\
    
    \includegraphics[width=0.21\textwidth]{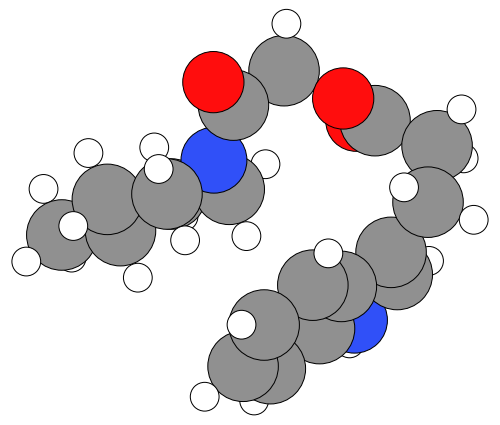} &
    \includegraphics[width=0.21\textwidth]{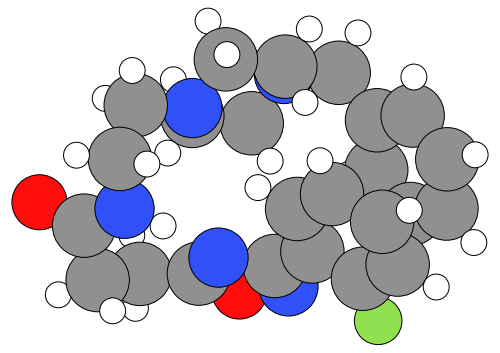} &
    \includegraphics[width=0.21\textwidth]{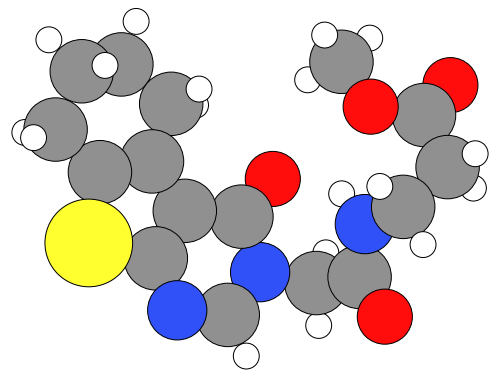} &
    \includegraphics[width=0.21\textwidth]{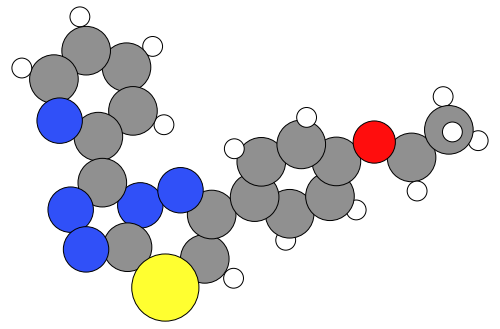}
    \\

    \includegraphics[width=0.21\textwidth]{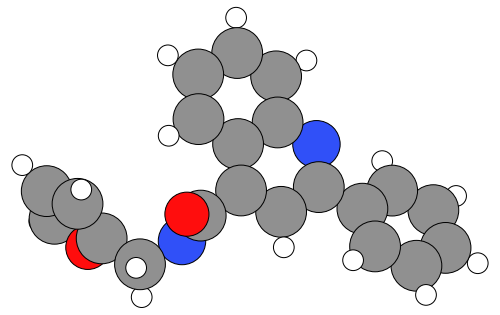} &
    \includegraphics[width=0.21\textwidth]{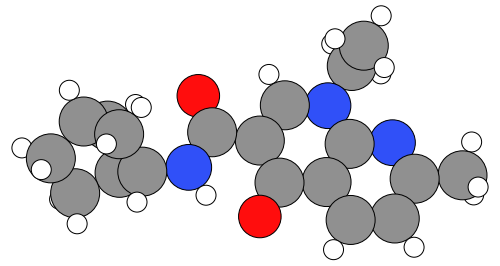} &
    \includegraphics[width=0.21\textwidth]{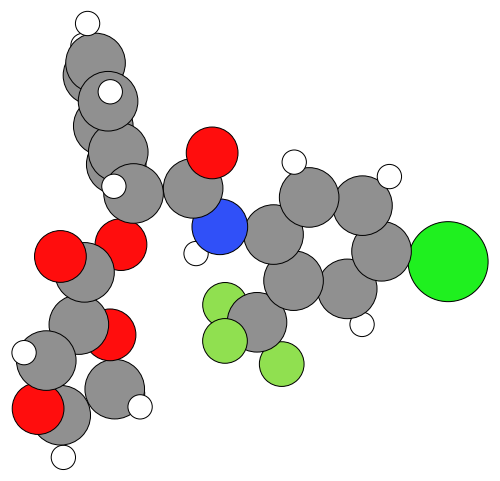} &
    \includegraphics[width=0.21\textwidth]{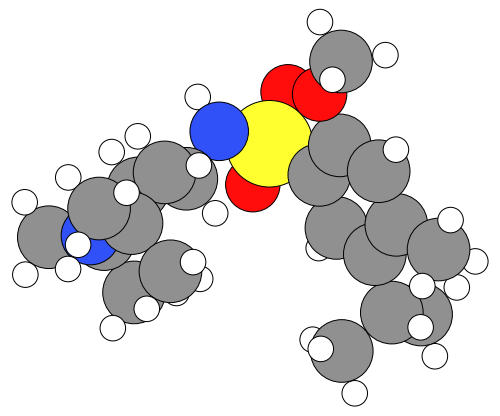}
    \\
    
    \includegraphics[width=0.21\textwidth]{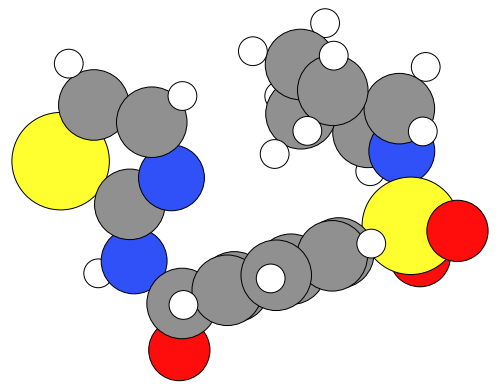} &
    \includegraphics[width=0.21\textwidth]{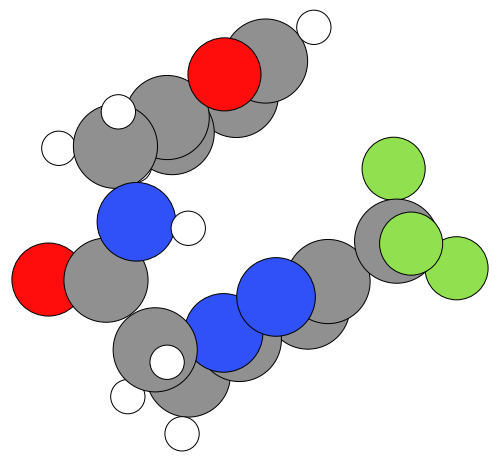} &
    \includegraphics[width=0.21\textwidth]{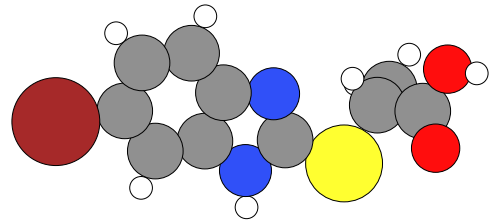} &
    \includegraphics[width=0.21\textwidth]{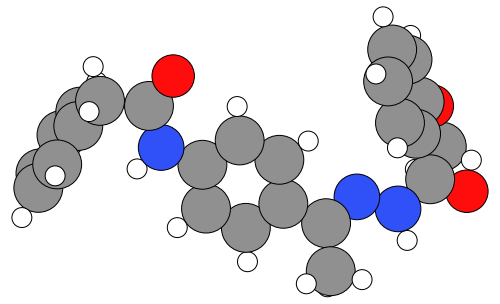}
    \\
    \end{tabular}
    \caption{Samples from \drugs{}}
    \label{fig:geom_samples}
\end{figure}

\newpage

\section{Figure \ref{fig:speedup} Outlier Structure}
\label{app:outlier}

\begin{figure}[h]
    \centering
    \begin{tblr}{cQ[c,m]c}
        \raisebox{-2.5cm}{\includegraphics[width=0.25\textwidth]{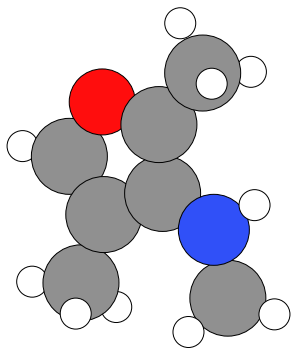}}
        &
        $\xrightarrow{\hspace*{30pt}}$
        &
        \raisebox{-2.5cm}{\includegraphics[width=0.25\textwidth]{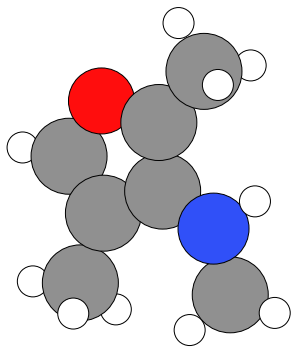}}
        \\
        MMFF & & Diffused \\
        $E=0$ kcal/mol  & & $E=-3.0$ kcal/mol 
        \\ 
        $\Bigg\downarrow$ & & $\Bigg\downarrow$ \\
        \includegraphics[width=0.25\textwidth]{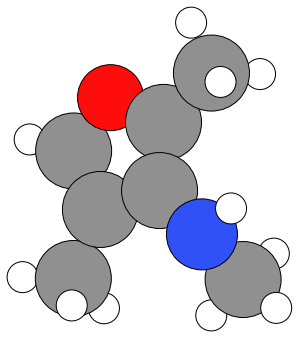} &
        \hspace{50pt} & 
        \includegraphics[width=0.25\textwidth]{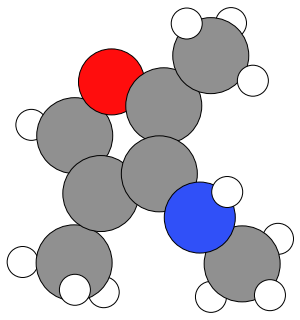}
        \\
        Relaxed MMFF & & Relaxed Diffused \\
        $E=-4.5$ kcal/mol & & $E=-5.4$ kcal/mol
    \end{tblr}
    \caption{Example molecule where the DFT-relaxed structures differed by $\sim0.9$kcal/mol when starting at the MMFF structure vs. the diffused structure.
    Diffusion rotates the hydrogens on the top methyl group slightly, which leads the subsequent DFT relaxation to find a better local minimum of the PES.
    Arrows between structures indicate which structure was used to initialize each calculation.
    Energies indicated are relative to the MMFF-relaxed structure and were computed using single-point DFT calculations.}
    \label{fig:speedup_outlier}
\end{figure}

\newpage
\section{Speedup Excluding Large Energy Gaps}
\label{app:lowE}
\begin{figure}[h]
    \centering
    \includegraphics[width=0.49\textwidth]{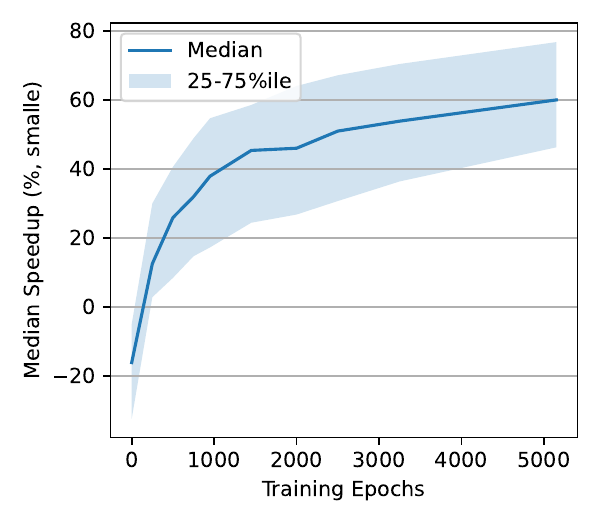}
    \includegraphics[width=0.49\textwidth]{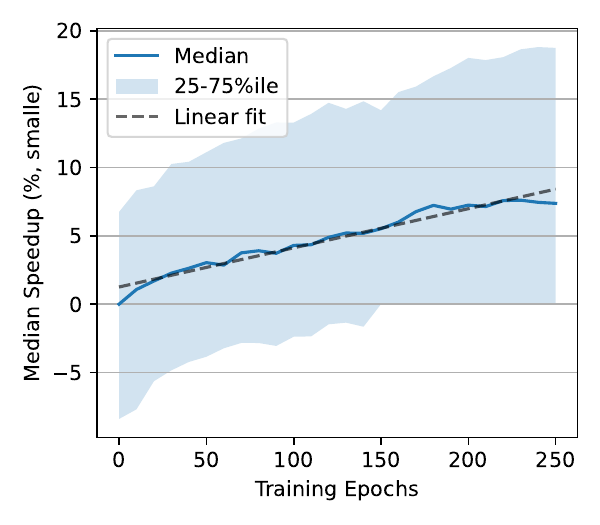}
    \caption{Speedup in number of DFT geometry steps required to converge as a function of how many epochs the model was trained for. Left: EDM trained on QM9. Right: EDM trained on \drugs{}. Structures were excluded from this plot if the absolute energy gap is greater than $0.2$ kcal/mol.}
    \label{fig:speedup_lowE}
\end{figure}

Figures \ref{fig:speedup} and \ref{fig:speedup_geom} show how the speedup in DFT relaxation time increases as the model is trained.
Because some structures have a nonzero energy gap, it is possible that some speedup in DFT results from diffusion finding structures that lie within shallower basins of the DFT PES.
To investigate this possibility, we reproduce the right-hand side of these figures when excluding any structure for which the absolute value of the energy gap is greater than $0.2$kcal/mol.
Results are shown in Figure \ref{fig:speedup_lowE}.

\end{document}